\documentclass[graybox]{svmult}

\usepackage{mathptmx}       
\usepackage{helvet}         
\usepackage{courier}        
\usepackage{type1cm}        
\usepackage{makeidx}         
\usepackage{graphicx}        
\usepackage{multicol}        
\usepackage[bottom]{footmisc}
\usepackage{float}
\usepackage[caption = false]{subfig}

\makeindex             

\begin{document}

\title*{Third and fifth harmonic responses in viscous liquids}
\author{S. Albert, M. Michl, P. Lunkenheimer, A. Loidl, P. M. D\'ejardin, and F. Ladieu }
\institute{S. Albert \at SPEC, CEA, CNRS, Universit\'e Paris-Saclay, CEA Saclay Bat 772, 91191 Gif-sur-Yvette Cedex, France. \email{samuel.albert@ens-lyon.fr}
\and M. Michl \at Experimental Physics V, Center for Electronic Correlations and Magnetism, University of Augsburg, 86159 Augsburg, Germany. \email{marion.michl@physik.uni-augsburg.de}
\and P. Lunkenheimer \at Experimental Physics V, Center for Electronic Correlations and Magnetism, University of Augsburg, 86159 Augsburg, Germany. \email{peter.lunkenheimer@physik.uni-augsburg.de}
\and A. Loidl \at Experimental Physics V, Center for Electronic Correlations and Magnetism, University of Augsburg, 86159 Augsburg, Germany. \email{alois.loidl@physik.uni-augsburg.de}
\and P. M. D\'ejardin \at LAMPS Universit\'e de Perpignan Via Domitia - 52 avenue Paul Alduy - 66860 Perpignan Cedex, France. \email{dejardip@univ-perp.fr}
\and F. Ladieu \at SPEC, CEA, CNRS, Universit\'e Paris-Saclay, CEA Saclay Bat 772, 91191 Gif-sur-Yvette Cedex, France. \email{francois.ladieu@cea.fr}}
%
%
\maketitle

\abstract*{We review the works devoted to third and fifth
harmonic susceptibilities in glasses, namely $\chi_3^{(3)}$ and
$\chi_{5}^{(5)}$. We explain why these nonlinear responses
are especially well adapted to test whether or not some
amorphous correlations develop upon cooling. We show that the
experimental frequency and temperature dependences
of  $\chi_3^{(3)}$ and of $\chi_5^{(5)}$ have anomalous features,
since their behavior is qualitatively different to that of an ideal
gas, which is the high temperature limit of a fluid.
Most of the works have interpreted this anomalous behavior as
reflecting    the growth, upon cooling, of amorphously ordered
 domains, as predicted by the general framework of Bouchaud and Biroli
  (BB). We explain why most -if not all- of the
  challenging interpretations can be recast in a way
  which is consistent with that of BB. Finally, the comparison
  of the anomalous features of $\chi_5^{(5)}$ and
  of $\chi_3^{(3)}$ shows that
  the amorphously ordered domains are compact,
  i.e. the fractal dimension $d_f$ is close to the dimension $d$
  of space. This suggests  that the glass transition of
  molecular liquids corresponds to a new universality class of
  critical phenomena.}

{



\section{Why measuring harmonic susceptibilities? Some facts and an oversimplified argument} \label{part1}
Most of our everyday materials are glasses, from window glasses to plastic bottles, and from colloids to pastes and granular materials. Yet the formation of the glassy state is still a conundrum
and the most basic questions about the nature of the glassy state remain unsolved, e.g., it is still hotly debated whether glasses are genuine solids or merely hyperviscous liquids.

Over the past three decades, the notion evolved that higher-order harmonic susceptibilities are especially well suited to unveil the very peculiar
correlations governing the glass formation, yielding information that cannot be accessed by monitoring the linear response. This is illustrated in Fig. \ref{fig1} displaying the third harmonic 
cubic susceptibility $\chi_3^{(3)}$ -defined in Section \ref{part2-1}- for four very different kinds of glasses \cite{Lev86,Hem96,Cra10,Bru11,Bra10,Sey16}. In the case of spin glasses \cite{Lev86,Bin86} -see Fig. \ref{fig1}(A)-, it was
discovered in the eighties that $\chi_3^{(3)}$ diverges at the spin glass transition temperature $T_{SG}$, revealing the long range nature of the spin glass amorphous order emerging
around $T_{SG}$. Here the expression ``amorphous order'' corresponds to a minimum of the free energy realized by a configuration which is not spatially periodic. Similar non-linear susceptibility experiments have been performed by Hemberger et al. \cite{Hem96} on an orientational glass former. In orientational glasses, electric dipolar or quadrupolar degrees of freedom undergo a cooperative freezing process without long-range orientational order \cite{Hoc90}. As illustrated in Fig. \ref{fig1}(B), the divergence of $\vert \chi_3^{(3)} \vert$ is not accompanied by any divergence of the linear susceptibility $\vert \chi_{1} \vert$.

\begin{figure}[t]
\includegraphics[width = 12cm]{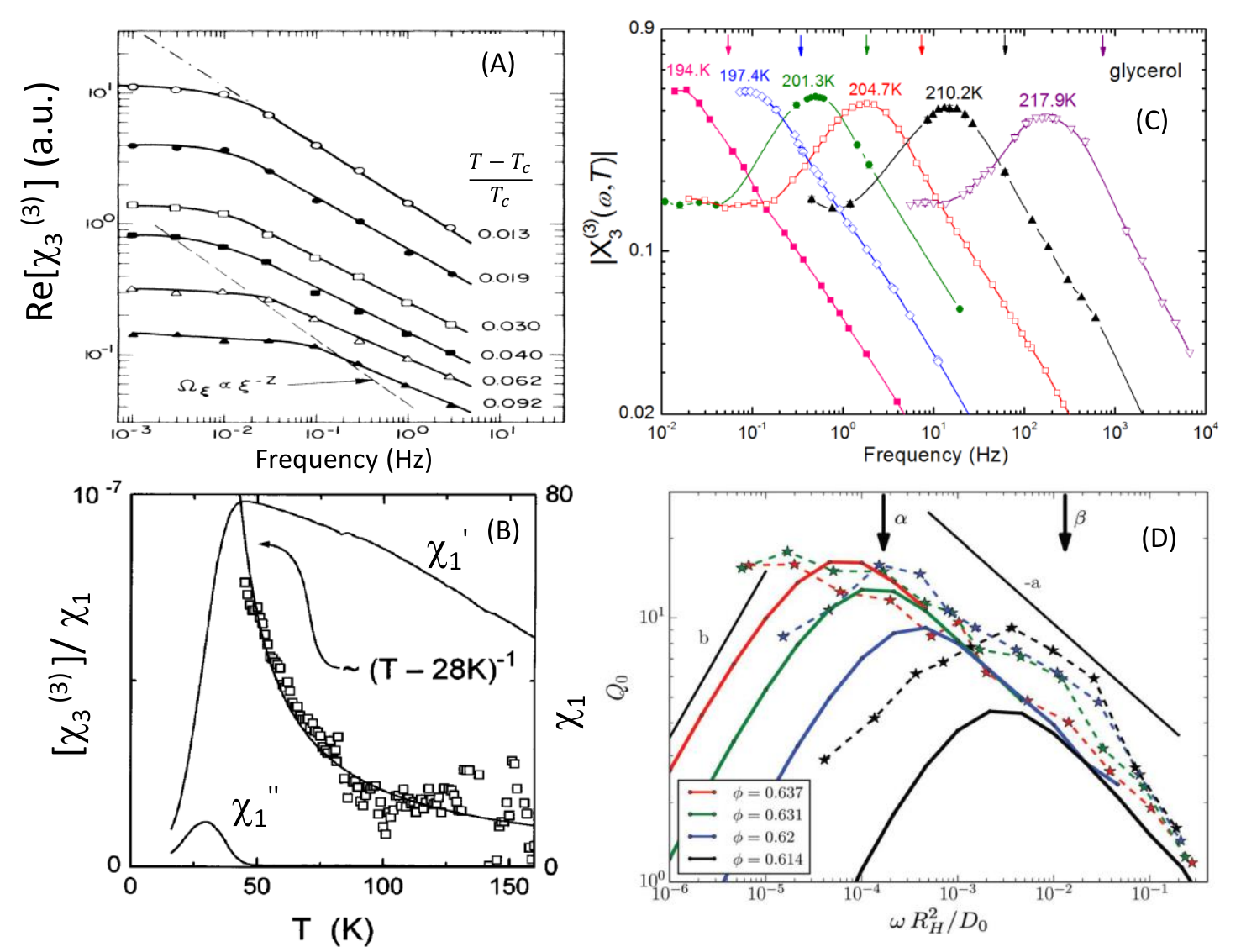}
\caption{From Refs. \cite{Lev86,Hem96,Cra10,Sey16}. Third-harmonic susceptibilities of very different types of glasses approaching their glass transition. \textbf{(A)}:
In the Ag:Mn spin glass \cite{Lev86}, the static value of $\chi_3^{(3)}$ diverges when approaching  the critical temperature $T_c \simeq 2.94$~K
 \cite{Lev86}. \textbf{(B)}: Similar arguments are used to rationalize the third-harmonic dielectric susceptibility of an orientational glass \cite{Hem96}. \textbf{(C)}: In glycerol \cite{Cra10,Bru11}, the modulus of the -dimensionless- cubic susceptibility $X_3^{(3)}$ has a peak
as function of frequency, which increases anomalously upon cooling. \textbf{(D)}: Strain-stress experiment in the colloidal system studied in Refs. \cite{Bra10,Sey16}.
When increasing the volumic density $\phi$ the increasing peak of $Q_0 = \vert \chi_3^{(3)}/ \chi_{1} \vert$ reveals that any shear strain connects the system
to a non equilibrium steady state -see \cite{Bra10,Sey16}-. In all these four examples $\chi_3^{(3)}$ unveils informations about the nature of the glassy
state that cannot be obtained by studying the linear susceptibility $\chi_{lin}$.}
\label{fig1}
\end{figure}

We shall show in Eqs. (\ref{eq1})-(\ref{eq2}) that this is intimately related to the very notion of amorphous ordering. For structural glasses,
e.g., glycerol, it was discovered \cite{Cra10,Bru11} less than $10$ years ago that $\vert \chi_3^{(3)}(\omega, T) \vert$ has a hump close to the $\alpha$ relaxation frequency $f_{\alpha}$, the height
of this hump increasing anomalously upon cooling. A hump of $\vert \chi_3^{(3)} \vert$ has also been recently discovered in a colloidal glass \cite{Bra10,Sey16}, in the vicinity of the
 $\beta$ relaxation frequency $f_{\beta}$, revealing that any shear strain connects the system to a non equilibrium steady state -see \cite{Bra10,Sey16}-. Of course, as detailed balance
 does not hold in colloids, the comparison of colloidal glasses with spin glasses, orientational glasses, and structural glasses cannot be quantitative. However, the four very different kinds of
 glasses of Fig. \ref{fig1} have the common qualitative property that nonlinear cubic responses unveil new information about the glassy state formation.

Let us now give an oversimplified argument explaining why nonlinear responses should unveil the correlations developing in glasses. We shall adopt the dielectric language adapted to
this review devoted to supercooled liquids -where detailed balance holds-, and consider a static electric field $E_{{st}}$ applied onto molecules carrying a dipole moment $\mu_{{dip}}$. At
high temperature $T$ the system behaves as an ideal gas and its  polarization $P$ is given by:

\begin{eqnarray}
P & = &
   \frac{\mu_{{dip}}}{a^d} {\cal L}_d \left( \frac{\mu_{{dip}}E_{st}}{k_BT} \right)  \nonumber \\
\ & \simeq &
  \frac{1}{3} \frac{\mu_{{dip}}}{a^d} \left( \frac{\mu_{{dip}}E_{st}}{k_BT} \right) -
 \frac{1}{45} \frac{\mu_{{dip}}}{a^d} \left( \frac{\mu_{{dip}}E_{st}}{k_BT} \right)^3 +
\frac{2}{945} \frac{\mu_{{dip}}}{a^d} \left( \frac{\mu_{{dip}}E_{st}}{k_BT} \right)^5 + ...
\label{eq1}
\end{eqnarray}
where $a^d$ is the molecular $d$-dimensional volume, ${\cal L}_d$ is the suitable Langevin function expressing the thermal equilibrium of a single dipole in dimension $d$, and where the
numerical prefactors of the linear, third, and fifth order responses correspond to the case $d=3$. Assume now that upon cooling some correlations develop over a characteristic lengthscale $\ell$, i.e.
molecules are correlated within groups containing $N_{{corr}} = (\ell/a)^{d_f}$ molecules, with $d_f$ the fractal dimension characterizing the correlated regions. Because these domains
are independent from each other, one can use Eq. (\ref{eq1}), provided that we change the elementary volume $a^d$ by that of a domain -namely $a^d (\ell/a)^d$-, as well as the molecular dipole
$\mu_{{dip}}$ by that of a domain -namely $\mu_{{dip}} (\ell /a)^{(d_f/2)}$-. Here, the exponent $d_f/2$ expresses the amorphous ordering within the correlated regions, i.e. the fact that the
orientation of the correlated molecules looks random in space. We obtain:
\begin{eqnarray}
\frac{P}{\mu_{{dip}}/a^d} & \simeq &
         \frac{1}{3} \left(\frac{\ell}{a}\right)^{d_f-d} \left( \frac{\mu_{{dip}}E_{st}}{k_BT} \right) -
         \frac{1}{45} \left(\frac{\ell}{a}\right)^{2d_f-d} \left( \frac{\mu_{{dip}}E_{st}}{k_BT} \right)^3 +  \nonumber \\
\ &\ & + \frac{2}{945} \left(\frac{\ell}{a}\right)^{3d_f-d} \left( \frac{\mu_{{dip}}E_{st}}{k_BT} \right)^5 + ...
\label{eq2}
\end{eqnarray}
which shows that the larger the order $k$ of the response, the stronger the increase of the response when $\ell$ increases.
As $d_f \le d$, Eq. (\ref{eq2}) shows that the linear
response \textit{never} diverges with $\ell$: it is always, for any $\ell$, of the order of $\mu_{{dip}}^2/(a^d k_B T)$. This can be seen directly in
Eq. (\ref{eq2}) in the case $d_f=d$
; while for $d_f<d$ one must add to  Eq. (\ref{eq2}) the polarization arising from the uncorrelated molecules not belonging to any correlated region. This
insensitivity of the linear response to $\ell$ directly comes from
the amorphous nature of orientations that we have assumed when rescaling the net dipole of a domain -by using the power $d_f/2$-. By contrast,
in a standard para-ferro transition one would
use instead a power $d_f$ to rescale the moment of a domain, and we would find that the linear response diverges with $\ell$ as soon as $d_f > d/2$ -which is the standard result
close to a second order phase transition-. For amorphous ordering, the cubic response is thus the lowest order response diverging with $\ell$, as soon as $d_f > d/2$. This is why cubic
responses -as well as higher order responses- are ideally suited to test whether or not amorphous order develops in supercooled liquids upon cooling.

For spin-glasses, the above purely thermodynamic argument is enough to relate the divergence of the static value of $\chi_3^{(3)}$
-see Fig \ref{fig1}-(A)- to the divergence of
the amorphous order correlation length $\ell$. For structural glasses this argument must be complemented by some dynamical argument, since we have seen on Fig. \ref{fig1}-(C)
that the anomalous behavior of $\chi_3^{(3)}$ takes place around the relaxation frequency $f_{\alpha}$. This has been done, on very general grounds, by the predictions of Bouchaud
 and Biroli (BB), who anticipated \cite{Bou05} the main features reported in Fig. \ref{fig1}-(C). BB's predictions will be explained in Section \ref{part3}. Before, we shall review in  Section \ref{part2} the main experimental features of third and fifth harmonic susceptibilities. Because of the generality of Eq. (\ref{eq2}) and of BB's framework,  we anticipate that $\chi_3$ and $\chi_5$ have common anomalous features that can be interpreted as reflecting the evolution of $\ell$ -and thus of $N_{corr}$- upon cooling.
 The end of the chapter, Section \ref{part4}, will be devoted to more specific approaches of the cubic response of glass forming liquids. Beyond their apparent diversity, we shall show that they can be unified by the fact that in all of them, $N_{corr}$ is a key parameter -even though it is sometimes implicit-. The Appendix contains
 some additional material for the readers aiming at deepening their understanding of this field of high harmonic responses.

\section{Experimental behavior of third and fifth harmonic  suceptibilities} \label{part2}
	\subsection{Definitions} \label{part2-1}
	
	When submitted to an electric field $E(t)$ depending on time $t$, the most general expression of the polarisation $P(t)$ of a dielectric medium is given by a series expansion :

\begin{equation}
P(t) = \sum_{m=0}^{\infty} P_{2m+1}(t)
\label{eq3}
\end{equation}
where because of the $E \to -E$ symmetry, the sum contains only odd terms, and the $(2m+1)$-order polarisation $P_{2m+1}(t)$ is proportional to $E^{2m+1}$. The most general expression of $P_{2m+1}(t)$ is given by:

\begin{equation}
\frac{P_{2m+1}(t)}{\epsilon_0} = \int_{-\infty}^{\infty}... \int_{-\infty}^{\infty} \chi_{2m+1}(t-t'_1, ...,t-t'_{2m+1}) E(t'_1)... E(t'_{2m+1}) dt'_1 ... dt'_{2m+1}
\label{eq4}
\end{equation}

\noindent Because of causality $ \chi_{2m+1} \equiv 0$ whenever one of its arguments is negative. For a field $E(t)=E\cos(\omega t)$ of frequency $\omega$ and of amplitude $E$,
it is convenient to replace $\chi_{2m+1}$ by its $(2m+1)$-fold Fourier transform and to integrate first over $t'_1, ..., t'_{2m+1}$. Defining the one-fold Fourier transform
 $\phi(\omega)$ of any function $\phi(t)$ by $\phi(\omega) = \int \phi(t) e^{-i\omega t}dt$ (with $i^2 = -1$) and using
 $\int e^{-i(\omega_1-\omega) t}dt = 2 \pi \delta(\omega_1-\omega)$, where $\delta$ is the Dirac delta function, one obtains the expression of $P_{2m+1}(t)$. This expression can be simplified
 by using two properties: \textit{(a)} the fact that the various frequencies $\omega_{\lambda}$ play the same role, which implies
 $\chi_{2m+1}(-\omega, \omega, ..., \omega) = \chi_{2m+1}(\omega, -\omega, ..., \omega)$; \textit{(b)} the fact that $\chi_{2m+1}$ is real in the time domain implying
 that $\chi_{2m+1}(-\omega,...,-\omega)$ is the complex conjugate of $\chi_{2m+1}(\omega,...,\omega)$. By using these two properties, we obtain the expression of all
 the $P_{2m+1}(t)$, and in the case of the third order polarisation this yields:

 \begin{equation}
\frac{P_{3}(t)}{\epsilon_0} = \frac{1}{4} E^3 \vert \chi_3^{(3)}(\omega) \vert \cos(3\omega t - \delta_3^{(3)}(\omega)) + \frac{3}{4} E^3 \vert \chi_3^{(1)}(\omega) \vert \cos(\omega t - \delta_3^{(1)}(\omega))
\label{eq5}
\end{equation}
where we have set $\chi_3(\omega,\omega, \omega) = \vert \chi_3^{(3)}(\omega) \vert e^{-i \delta_3^{(3)}(\omega)}$, and $\chi_3(\omega, \omega, -\omega) = \vert \chi_3^{(1)} (\omega) \vert e^{-i \delta_3^{(1)}(\omega)}$.

Similarly, for the fifth-order polarisation, we obtain:

\begin{eqnarray}
\frac{P_{5}(t)}{\epsilon_0} & = & \frac{1}{16} E^5 \vert \chi_5^{(5)}(\omega) \vert \cos(5\omega t - \delta_5^{(5)}(\omega)) +
 \frac{5}{16} E^5 \vert \chi_5^{(3)}(\omega) \vert \cos(3\omega t - \delta_5^{(3)}(\omega)) + \nonumber  \\
 \ & \ & +  \frac{10}{16} E^5 \vert \chi_5^{(1)}(\omega) \vert \cos(\omega t - \delta_5^{(1)}(\omega))
\label{eq6}
\end{eqnarray}
where, we have set $\chi_5(\omega,\omega, \omega, \omega, \omega) = \vert \chi_5^{(5)}(\omega) \vert e^{-i \delta_5^{(5)}(\omega)}$, and
similarly $\chi_5(\omega,\omega, \omega, \omega, -\omega) = \vert \chi_5^{(3)}(\omega) \vert e^{-i \delta_5^{(3)}(\omega)}$ as well
as $\chi_5(\omega,\omega, \omega, -\omega, -\omega) = \vert \chi_5^{(1)}(\omega) \vert e^{-i \delta_5^{(1)}(\omega)}$.

For completeness we recall that the expression of the linear polarisation $P_1(t)$ is $P_{1}(t)/{\epsilon_0} =  E \vert \chi_1(\omega) \vert \cos(\omega t - \delta_1(\omega))$
where we have set $\chi_1(\omega) = \vert \chi_1(\omega) \vert e^{-i \delta_1(\omega)}$. In the linear case, we often drop the exponent indicating the harmonic, since the linear response
$P_1(t)$ is by design  at the fundamental angular frequency $\omega$. The only exception to this simplification is in Fig. \ref{fig11} (see below)
where for convenience the linear susceptibility is denoted $\chi_1^{(1)}$.

Up to now we have only considered nonlinear responses induced by a pure ac field $E$, allowing to define  the third harmonic cubic susceptibility $\chi_3^{(3)}$ and/or the fifth-harmonic fifth-order susceptibility $\chi_5^{(5)}$ to which this chapter is devoted. In
Section \ref{part2-3} and Figs. \ref{fig8}-\ref{fig9}, we shall briefly compare $\chi_3^{(3)}$ with other cubic susceptibilities, namely $\chi_{3}^{(1)}$ already defined in Eq. (\ref{eq5}) as well as $\chi_{2;1}^{(1)}$ that we introduce now.

This supplementary cubic susceptibility is one of the new terms  arising when a static field $E_{{st}}$ is
superimposed on top of $E$. Because of $E_{st}$, new cubic responses arise, both for even and odd harmonics. For brevity, we shall write only the expression of the first harmonic part $P_{3}^{(1)}$ of the cubic polarization, which now contains two terms:

\begin{equation}
\frac{P_{3}^{(1)}(t)}{\epsilon_0} = \frac{3}{4} \vert \chi_{3}^{(1)} (\omega) \vert E^3\cos{(\omega t - \delta_{3}^{(1)}(\omega))} + 3 \vert \chi_{2,1}^{(1)}(\omega)\vert E_{{st}}^2 E\cos{(\omega t - \delta_{2,1}^{(1)}(\omega))}
\label{eq7}
\end{equation}
where we have defined $\vert \chi_{2,1}^{(1)}(\omega)\vert  \exp{(-i \delta_{2,1}^{(1)}(\omega))} = \chi_3(0,0,\omega)$.

For any cubic susceptibility -- generically noted $\chi_3$ -- or for any fifth-order susceptibility -- generically noted $\chi_5$ --
the corresponding dimensionless susceptibility $X_3$ or $X_5$ is defined as :
\begin{equation}
X_3 \equiv \frac{k_B T}{\epsilon_0 \Delta \chi_1^2 a^3} \chi_3, \qquad
X_5 \equiv \frac{(k_B T)^2 }{\epsilon_0^2 \Delta \chi_1^3 a^6} \chi_5
\label{eq8}
\end{equation}
where $\Delta \chi_1$ is the ``dielectric strength'', i.e. $\Delta \chi_1 = \chi_{{lin}}(0) - \chi_{{lin}}(\infty)$ where $\chi_{{lin}}(0)$ is the linear susceptibility at zero frequency and
$\chi_{{lin}}(\infty)$ is the linear susceptibility at a -high- frequency  where the orientational mechanism has ceased to operate. Note that $X_3$ as well as $X_5$ have the great advantage to be both
dimensionless and independent of the field amplitude.

	\subsection{Frequency and temperature dependence of third harmonic  susceptibility} \label{part2-2}
	
In this section we review the characteristic features of $\chi_{3}^{(3)}$ both as a function of frequency and temperature. We separate the effects at equilibrium above $T_g$ and those recorded below $T_g$ in the out-of-equilibrium regime.

\begin{center}
\begin{figure}[t]
\includegraphics[width = 10cm]{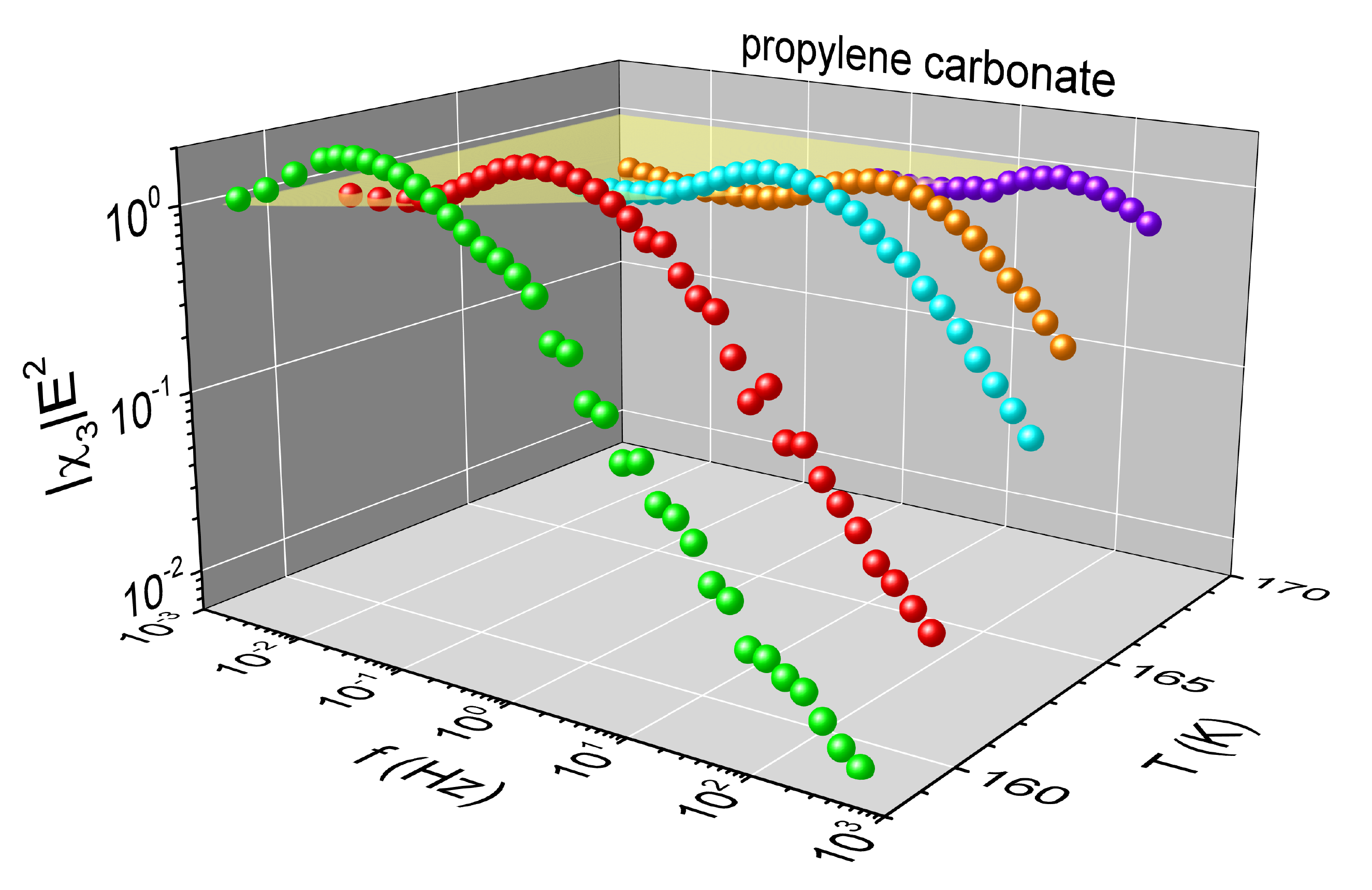}
\caption{Third-order harmonic component of the dielectric susceptibility of propylene carbonate \cite{Bau13}. Spectra of $\vert \chi_{3}^{(3)} \vert E^2 $ are shown for various temperatures measured at a field of 225~kV/cm. The yellow-shaded plane indicates the plateau arising in the trivial regime.}
\label{fig2}
\end{figure}
\end{center}	
	
		\subsubsection{Above $T_{g}$} \label{part2-2-1}

			\paragraph{In the $\alpha$ regime:}
			
Fig. \ref{fig2} shows the modulus $\vert \chi_{3}^{(3)} \vert$ for propylene carbonate \cite{Bau13}. It is an archetypical example of what has been measured in glass forming liquids close to $T_g$. For a given temperature one distinguishes two domains:

\begin{enumerate}

	\item For very low frequencies, $f/f_{\alpha} \le 0.05$, a plateau is observed as indicated by the shaded area in Fig. \ref{fig2}, i.e. $\vert \chi_3^{(3)} \vert$ does not depend on frequency. This
is reminiscent of the behavior of an ideal gas of dipoles where each dipole experiences a Brownian motion without any correlation with other dipoles. In
such an ideal gas, $\vert \chi_3^{(3)}\vert$ has a plateau below the relaxation frequency and monotonously falls to zero as one increases the frequency. Because 
the observed plateau in Fig. \ref{fig2} is reminiscent to the ideal gas case, it has sometimes \cite{Cra10,Bru11} been  called the ``trivial''
regime. What is meant here is not that the analytical expressions of the various $\chi_{3}$ are ``simple'' --see Section \ref{part6}--,
but that the glassy correlations do not change qualitatively the shape of $\chi_{3}^{(3)}$ in this range. Physically, an ideal gas of dipoles corresponds  to the high-$T$ limit of a fluid. This is why it is a useful benchmark which allows to distinguish the ``trivial'' features and those involving glassy correlations.

	\item When rising the frequency above $0.05 f_{\alpha}$ one observes for $\vert \chi_3^{(3)} \vert$  a hump for a frequency
$f_{peak}/f_{\alpha} \simeq c$ where the constant $c$ does not depend on $T$ and weakly depends on the liquid (e.g., $c \simeq 0.22$ for glycerol and $c \simeq 0.3$ for propylene carbonate). This hump is followed by a power law decrease $\vert \chi_3^{(3)} \vert \sim  f^{-\beta_3}$ where $\beta_3 <1$ is close \cite{Cra10} to the
exponent governing the decrease of $\vert \chi_{1} \vert$ above $f_{\alpha}$. Qualitatively, this hump is
important since it exists \textit{neither} in the cubic susceptibility of an ideal gas of dipoles \textit{nor} in the modulus
of the linear response $\vert \chi_{1} \vert$ of the supercooled liquids. This is why this hump has been termed the ``glassy contribution'' to $\chi_3$.
On a more quantitative basis, the proportionality of $f_{peak}$ and of $f_{\alpha}$ has been observed for $f_{\alpha}$ ranging from $0.01$~Hz to $10$~kHz
-above $10$~kHz the measurement of $\chi_3^{(3)}$ is obscured by heating issues, see \cite{Bru10} and Section \ref{part6}-.
	
\end{enumerate}

\begin{figure}[h]
\includegraphics[width = 8cm]{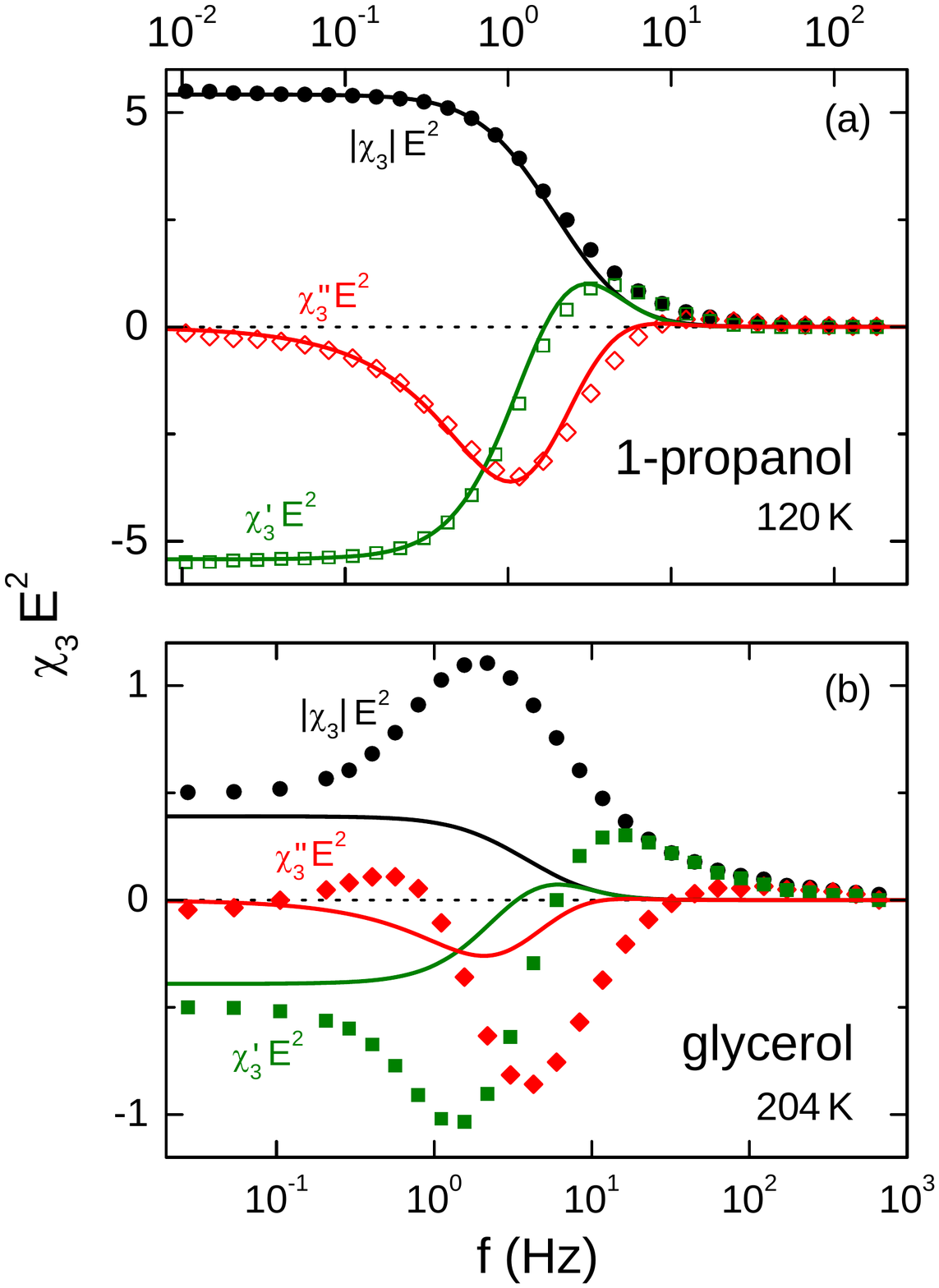}
\caption{(a) Modulus, real, and imaginary part of the third-order dielectric susceptibility $\chi_{3}^{(3)}$ (times $E^{2}$) of 1-propanol at 120~K as measured with a field of 468~kV/cm \cite{Lun17}. The solid lines were calculated according to Refs. \cite{Cof76}. (b) Same for glycerol at 204~K and 354~kV/cm \cite{Lun17}.
}
\label{fig3}
\end{figure}

The consistency of the above considerations can be checked by comparing the third-order susceptibility of canonical glass formers to that of monohydroxy alcohols. The linear dielectric response of the latter is often dominated by a Debye relaxation process, which is commonly ascribed to the fact that part of the molecules are forming chain-like hydrogen-bonded molecule clusters with relatively high dipolar moments \cite{Gai10}. This process represents an idealised Debye-relaxation case as it lacks the heterogeneity-related broadening found for other glass formers. Moreover, correlations or cooperativity should not play a significant role for this process, because cluster-cluster interactions can be expected to be rare compared to the intermolecular interactions governing the $\alpha$ relaxation in most canonical glass formers \cite{Bau15}. Thus, this relaxation process arising from rather isolated dipolar clusters distributed in a liquid matrix can be expected to represent a good approximation of the "ideal dipole gas" case mentioned above. The monohydroxy alcohol 1-propanol is especially well suited to check this notion because here transitions between different chain topologies, as found in several other alcohols affecting the nonlinear response \cite{Sin12,Sin13}, do not seem to play a role \cite{Sin13}. Figure \ref{fig3}(a) shows the frequency-dependent modulus, real, and imaginary part of $\chi_{3}^{(3)}E^2$ for 1-propanol at 120~K \cite{Bau15,Lun17}. Indeed, no hump is observed in $\vert\chi_{3}^{(3)}\vert(\nu)$ as predicted for a non-cooperative Debye relaxation. The solid lines were calculated according to Refs. \cite{Cof76}, accounting for the expected trivial polarization-saturation effect. Indeed, the spectra of all three quantities are reasonably described in this way. In the calculation, for the molecular volume an additional factor of 2.9 had to be applied to match the experimental data, which is well consistent with the notion that the Debye relaxation in the monohydroxy alcohols arises from the dynamics of clusters formed by several molecules.

In marked contrast to this dipole-gas-like behavior of the Debye relaxation of 1-propanol, the $\chi_{3}^{(3)}$ spectra related to the conventional $\alpha$ relaxation of canonical glass formers exhibit strong deviations from the trivial response, just as expected in the presence of molecular correlations. As an example, Fig. \ref{fig3}(b) shows the modulus, real, and imaginary part of $\chi_{3}^{(3)}E^2$ of glycerol at 204~K. Again the lines were calculated assuming the trivial nonlinear saturation effect only \cite{Cof76}. Obviously, this approach is insufficient to provide a reasonable description of the experimental data. Only the detection of plateaus in the spectra arising at low frequencies agrees with the calculated trivial response. This mirrors the fact that, on long time scales, the liquid flow smoothes out any glassy correlations.

When varying the temperature, two very different behaviors of $\chi_{3}^{(3)}$ are observed:

\begin{enumerate}

\item In the plateau region the weak temperature dependence of $\chi_{3}^{(3)}$ is easily captured by converting $\chi_3^{(3)}$  into
its dimensionless form $X_{3}^{(3)}$ by using Eq. (\ref{eq8}): one observes \cite{Cra10,Bru11} that \textit{in the plateau region} $X_{3}^{(3)}$
\textit{does not depend at all on the temperature}. Qualitatively this is important since in an ideal gas of dipoles $X_{3}^{(3)}$ does also not depend on temperature, once plotted as a function of $f/f_{\alpha}$. This reinforces the ``trivial'' nature of the plateau region, i.e. the fact that it is not
qualitatively affected by glassy correlations.

\begin{figure}[h]
\includegraphics[width = 8cm]{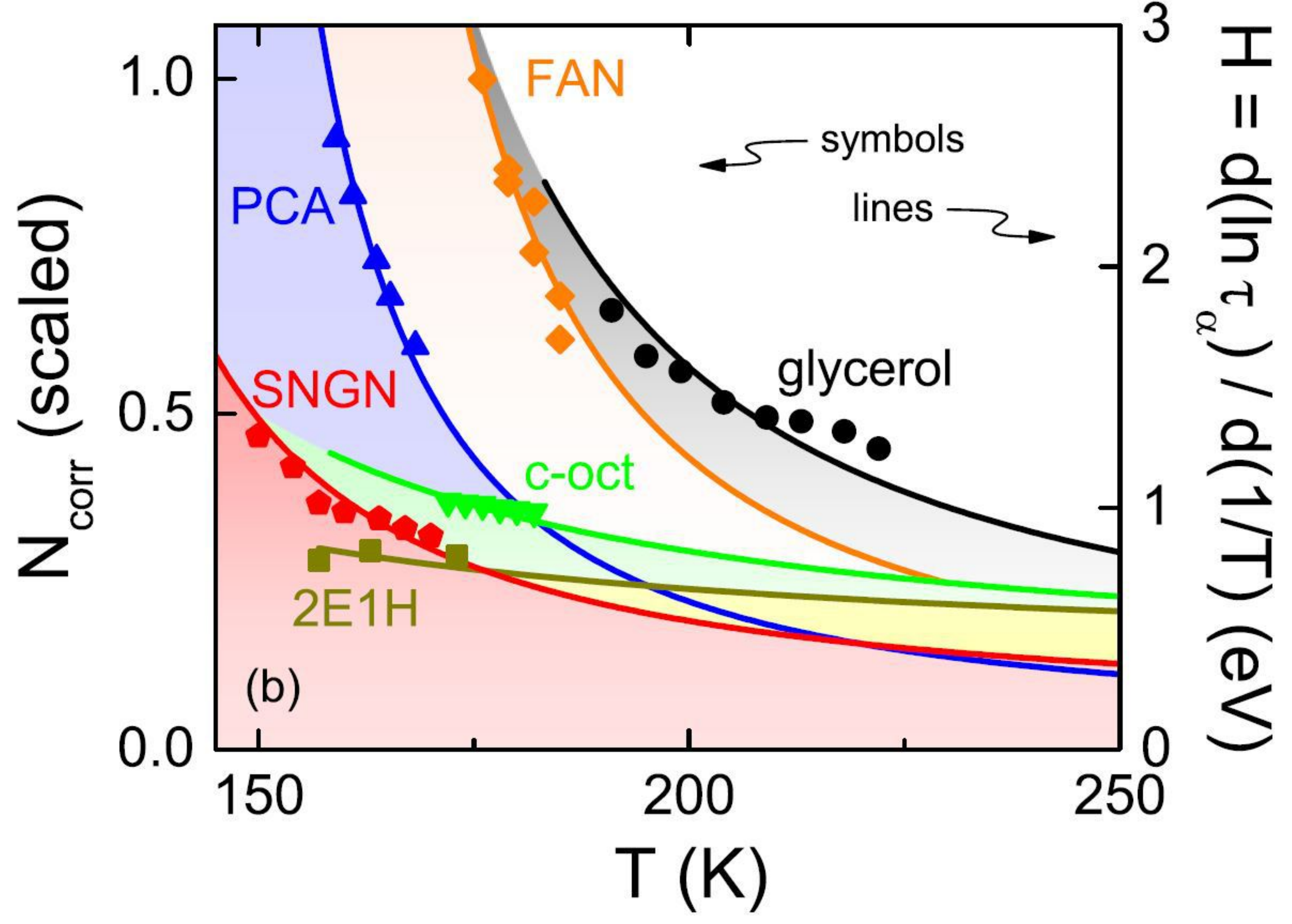}
\caption{From Ref. \cite{Mic16}: For several glass formers, $N_{corr}(T)$ as extracted from the hump of $\vert X_3^{(3)} \vert$ (left axis) closely follows $E_{act}(T)$, deduced from the temperature dependence of the $\alpha$-relaxation time \cite{Bau13} (right axis). The abbreviations stand for propylene carbonate (PCA), 3-fluoroaniline
(FAN), 2-ethyl-1-hexanol (2E1H), cyclo-octanol (c-oct), and a mixture of $60\%$ succinonitrile and $40\%$ glutaronitrile (SNGN).}
\label{fig4}
\end{figure}

\item In the hump region, $\vert X_{3}^{(3)} (f/f_{\alpha}) \vert$ increases upon cooling, again emphasizing the ``anomalous'' --or ``non trivial''-- behavior of
the glassy contribution to $\chi_{3}^{(3)}$. This increase of the hump of $\vert X_{3}^{(3)} \vert$ has been related to that of the apparent activation
energy $E_{act}(T) \equiv \partial \ln \tau_{\alpha} / \partial (1/T)$ -see refs. \cite{Bau13,Mic16}- as well as to
$T \chi_T \equiv \vert \partial \ln \tau_{\alpha} / \partial \ln T \vert$ \cite{Cra10,Bru11,Lho14,Cas15}. Note that because the experimental temperature interval is not
so large, the temperature behavior of $E_{act}$ and of $T \chi_T$ is extremely similar. Both quantities are physically appealing since they are related to the
number $N_{corr}(T)$ of correlated molecules: the line of thought where $E_{act} \sim N_{corr}(T)$ dates back to the work of Adam and Gibbs \cite{Ada65}; while
another series of papers \cite{Ber05,Dal07} proposed a decade ago that $N_{corr} \propto T \chi_T$. Fig. \ref{fig4} illustrates how good is the correlation
between the increase of the hump of $\vert X_3^{(3)} \vert$ -left axis- and $E_{act}(T)$. This correlation holds for $5$ glass formers, of
extremely different fragilities, including a plastic crystal, where only the orientational degrees of freedom experience the glass transition \cite{Bra02}.

\end{enumerate}

			\paragraph{In the excess wing regime:}

In the dielectric-loss spectra of various glass formers, at high frequencies the excess wing shows up, corresponding to a second, shallower  power law at the right flank of the $\alpha$ peak \cite{Lun00}. Figure \ref{fig5}(a) shows loss spectra of glycerol, measured at low and high fields up to 671~kV/cm \cite{Sch98,Bau13a}, where the excess wing is indicated by the dashed lines. (It should be noted that the difference of these loss curves for high and low fields is directly related to the cubic susceptibility $\chi_{3}^{(1)}$, defined in Eq. (\ref{eq5}) \cite{Lun17}.) As already reported in the seminal paper by Richert and Weinstein \cite{Ric06}, in Fig. \ref{fig5}(a) at the right flank of the $\alpha$-relaxation peak a strong field-induced increase of the dielectric loss is found while no significant field dependence is detected at its low-frequency flank. In Ref. \cite{Ric06} it was pointed out  that these findings are well consistent with the heterogeneity-based box model (see Section \ref{part4-3}). However, as revealed by Fig. \ref{fig5}(a), remarkably in the region of the excess wing no significant nonlinear effect is detected. Time-resolved measurements, later on reported by Samanta and Richert \cite{Sam14}, revealed nonlinearity effects in the excess-wing region when applying the high field for extended times of up to several 10000 cycles. Anyhow, the nonlinearity in this region seems to be clearly weaker than for the main relaxation and the nonlinear behavior of the excess wing differs from that of the $\alpha$ relaxation.

\begin{figure}[h]
\includegraphics[width = 9cm]{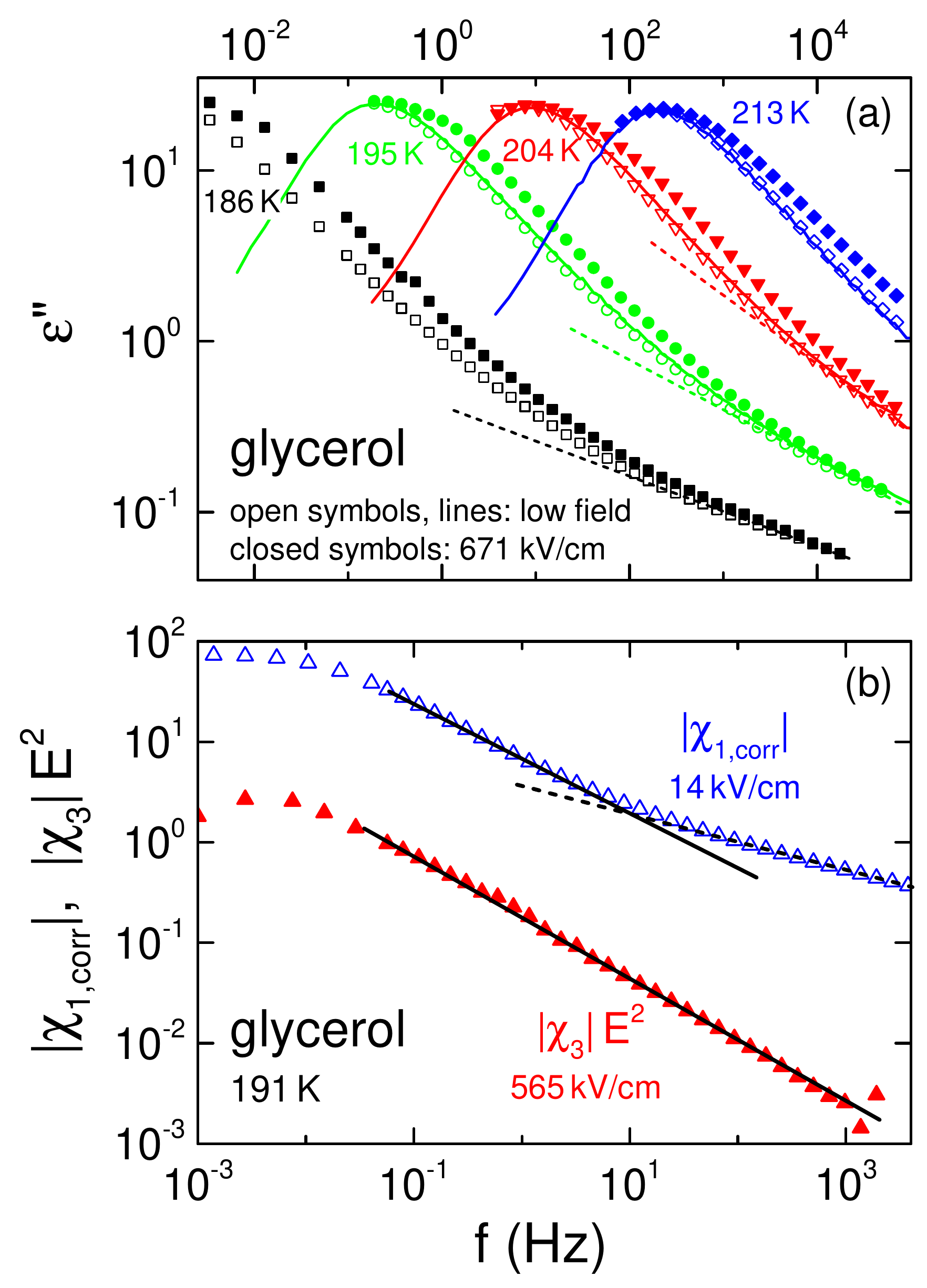}
\caption{(a) Dielectric loss of glycerol measured at fields of 14~kV/cm (open symbols) and 671~kV/cm (closed
symbols) shown for four temperatures \cite{Bau13a}. The solid lines were measured with 0.2~kV/cm \cite{Sch98}. The dashed lines indicate the excess wing. (b) Open triangles: Absolute values of $\chi_1$ (corrected for $\chi_{1,\infty}=\epsilon_{\infty}-1$) at 14~kV/cm for glycerol at 191~K. Closed triangles: $\chi_3^{(3)} E^2$ at 565~kV/cm \cite{Bau14}. The solid lines indicate similar power laws above the peak frequency for both quantities. The dashed line indicates the excess wing in the linear susceptibility at high frequencies, which has no corresponding feature in $\chi_3^{(3)}(\nu)$.}
\label{fig5}
\end{figure}

To check whether weaker nonlinearity in the excess-wing region is also revealed in higher-harmonic susceptibility measurements, Fig. \ref{fig5}(b) directly compares the modulus of the linear dielectric susceptibility of glycerol at 191~K to the third-order susceptibility $\vert\chi_{3}^{(3)}\vert$ (multiplied by $E^{2}$) \cite{Bau14}. (We show $\vert\chi_1\vert$ corrected for $\chi_{1,\infty}=\epsilon_{\infty}-1$ caused by the ionic and electronic polarizability, whose contribution in the modulus strongly superimposes the excess wing.) While the linear response exhibits a clear signature of the excess wing above about 100~Hz (dashed line), no trace of this spectral feature is found in $\vert\chi_{3}^{(3)}(\nu)\vert$. Thus, we conclude that possible nonlinearity contributions arising from the excess wing, if present at all, must be significantly weaker than the known power-law decay of the third-order susceptibility at high frequencies, ascribed to the nonlinearity of the $\alpha$ relaxation.

The excess wing is often regarded as the manifestation of a secondary relaxation process, partly superimposed by the dominating $\alpha$-relaxation \cite{Sch00,Dos02}. Thus the weaker nonlinearity of the excess wing seems to support long-standing assumptions of the absence of cooperativity in the molecular motions that lead to secondary relaxation processes \cite{Bei01,Nga03}. Moreover, in a recent work \cite{Nga15} it was pointed out that the small or even absent nonlinear effects in the excess-wing region can also be consistently explained within the framework of the coupling model \cite{Nga03}, where the excess wing is identified with the so-called "nearly constant loss" caused by caged molecular motions.

		\subsubsection{Below $T_{g}$} \label{part2-2-2}

\begin{figure}[t]
\includegraphics[width = 8cm]{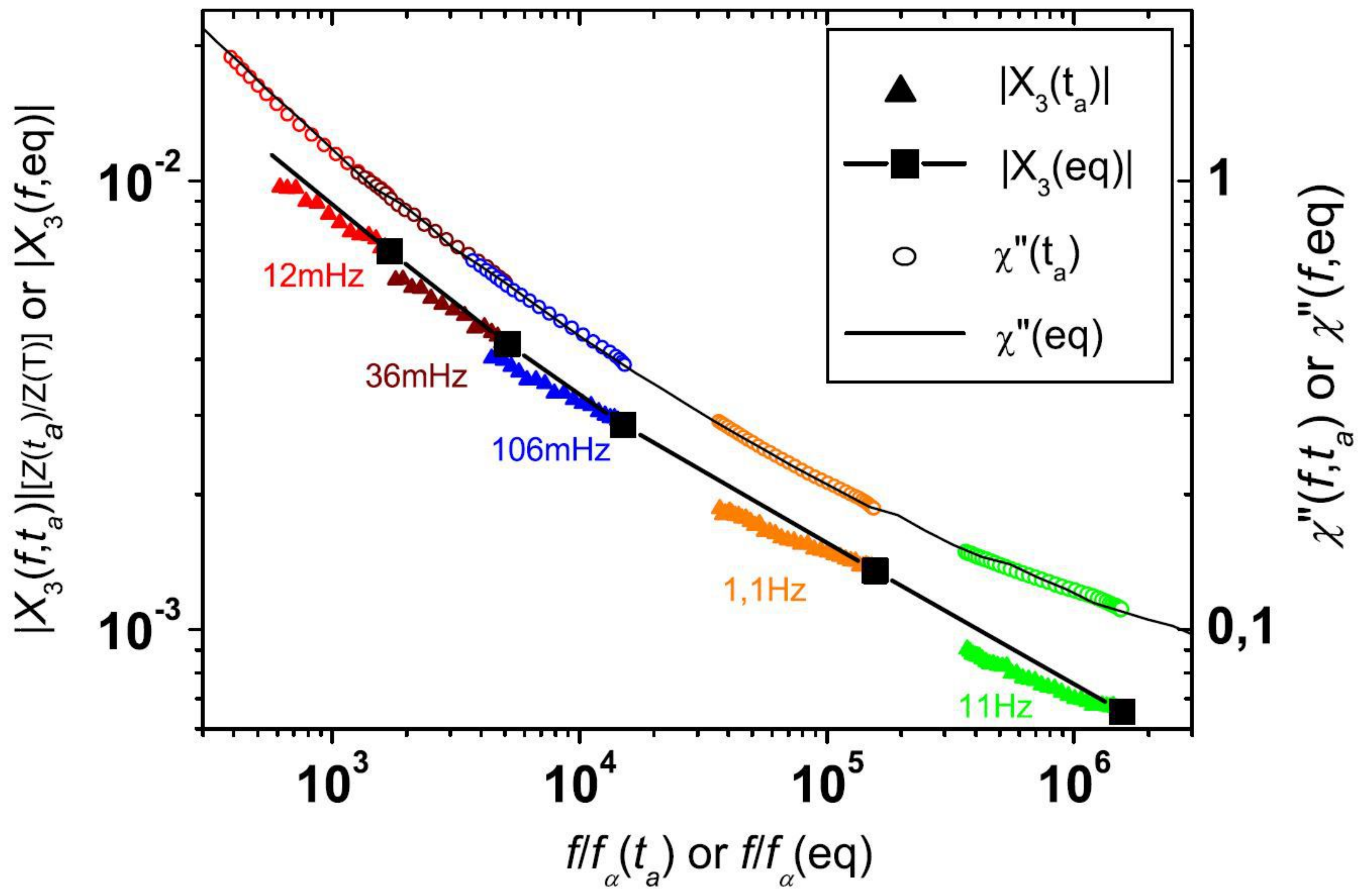}
\caption{From \cite{Bru12}. During the aging of glycerol -at $T_g-8$K- the increase of $\tau_{\alpha}$ with the aging time $t_a$ is measured by rescaling the aging data -symbols- of $\chi''_1$ -right axis- onto the equilibrium data -solid black line-. The corresponding scaling \textit{fails} for $X_3^{(3)}(f,t_a)$  -left axis- revealing the increase of $N_{corr}$ during aging. See \cite{Bru12} for details about the quantity $z(t_a)/z(T)$ which is involved in the left axis but varies by less than $2\%$ during aging-.}
\label{fig6}
\end{figure}

\begin{figure}[h]
\includegraphics[width = 8cm]{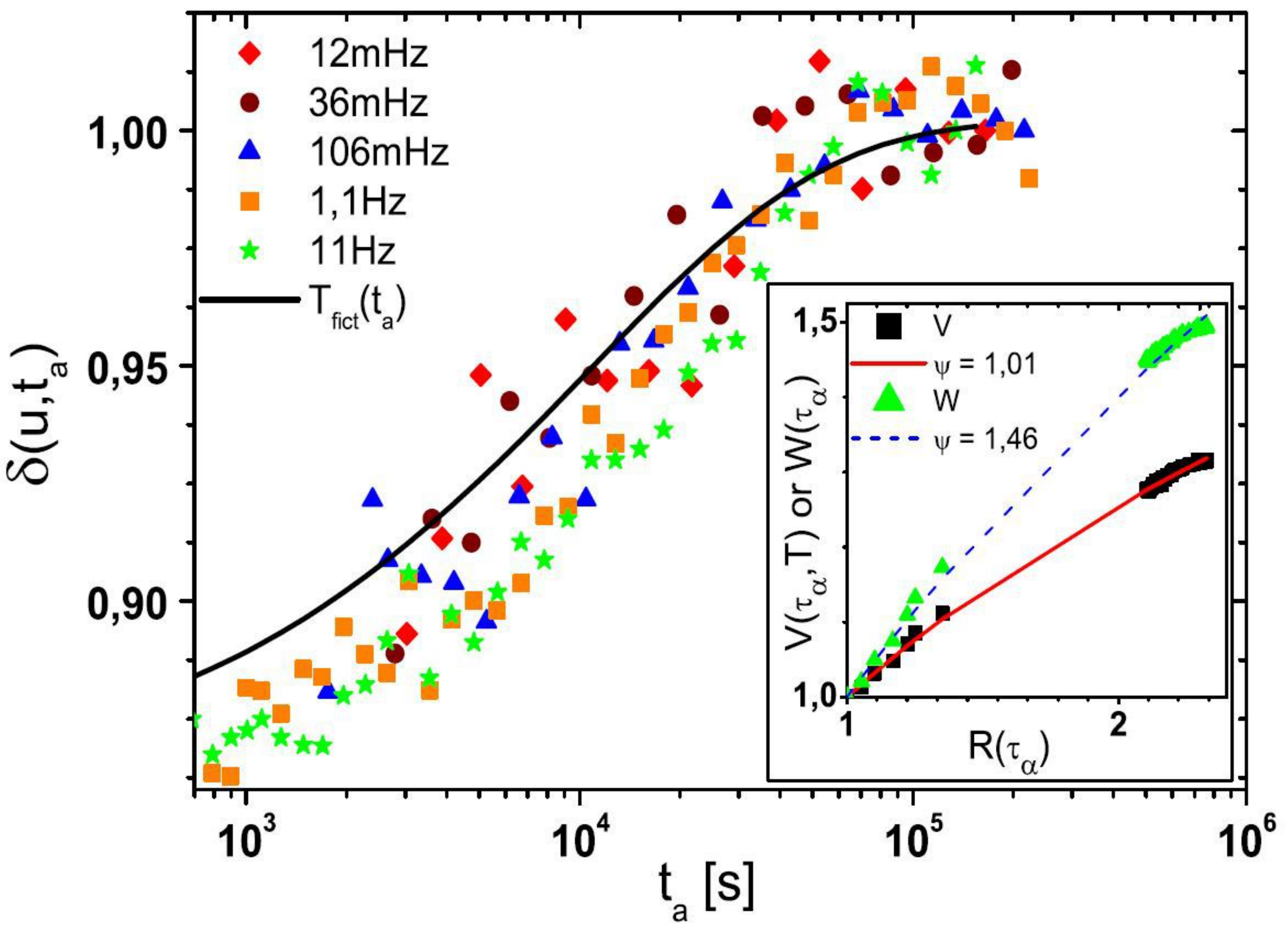}
\caption{From \cite{Bru12}. The values of $\delta = N_{corr}(t_a)/N_{corr}(eq)$ extracted from Fig. \ref{fig6} show the increase of $N_{corr}$ during aging. Inset: different theories are tested gathering equilibrium and aging experiments.}
\label{fig7}
\end{figure}

Below $T_g$, the physical properties are \textit{aging}, i.e. they depend on the time $t_a$ elapsed since the material has fallen out of equilibrium,
i.e. since the glass transition temperature $T_g$ has been crossed. The mechanism of aging is still a matter of controversy
\cite{Ber11,Too46,Nar71,Moy76,Lub04}, owing to the enormous theoretical and numerical difficulties inherent to out-of-equilibrium processes. Experimentally,
a few clear cut results have been obtained in spin glasses \cite{Ber04} where it was shown, by using nonlinear techniques, that the increase of the
relaxation time $\tau_{\alpha}$ with the aging time $t_a$ can be rather convincingly attributed to the growth of
 the number $N_{corr}$ of correlated spins with $t_a$. Very recently extremely sophisticated numerical simulations have been carried out by the so called Janus
international collaboration, yielding, among many other results, a strong microscopic support \cite{Janus} to the interpretation given previously in the experiments of
Ref. \cite{Ber04}.

In structural glasses, the aging properties of the linear response have been reported more than one decade ago
\cite{Leh98,Lun05}. More recently, the aging properties of $\chi_3^{(3)}$ were reported in glycerol \cite{Bru12} and its main outputs are summarized
in Figs. \ref{fig6} and \ref{fig7}. A glycerol sample previously well equilibrated at $T_g+8$~K was quenched to a working temperature $T_w = T_g-8$~K
and its third harmonic cubic susceptibility was continuously monitored as a function of $t_a$. The dominant effect is the increase of the relaxation
time $\tau_{\alpha}$ with $t_a$. In Ref. \cite{Bru12} $\tau_{\alpha}$ increases by a factor $\simeq 6$ between the arrival
at $T_w$ -i.e. $t_a = 0$- and the finally equilibrated situation reached for $t_a \gg \tau_{\alpha, eq}$ where $\tau_{\alpha}$ is equal to
its equilibrium value $\tau_{\alpha,eq}$ -and no longer evolves with $t_a$-. This variation of $\tau_{\alpha}$ with the aging time $t_a$
can be very accurately deduced from the shift that it produces on the imaginary part of the linear response $\chi''(f,t_a)$. This is summarized
in Fig. \ref{fig6} for $5$ different frequencies: when plotted as a function of $f/f_{\alpha}(t_a) \equiv 2 \pi f \tau_{\alpha}(t_a)$, the
aging values of $\chi''(f,t_a)$ -symbols- are nicely rescaled onto the equilibrium values $\chi''(f,eq)$ -continuous line-
measured when $t_a \gg \tau_{\alpha,eq}$. The most important experimental result is that this scaling \textit{fails} for
$\vert X_3^{(3)}(f,t_a)\vert$ as shown by the left axis of Fig. \ref{fig6}: For short aging times, the difference between aging data (symbols) and equilibrium values (continuous line) is largest.
This has been interpreted as \textit{an increase of $N_{corr}$ with the aging time} $t_a$. This increase of $N_{corr}(t_a)$ towards its equilibrated value $N_{corr}(eq)$ is illustrated in Fig. \ref{fig7} where the variation of $\delta = N_{corr}(t_a)/N_{corr}(eq)$ is plotted as a function of $t_a$. It turns out to be independent of the measuring frequency, which is a very important
self consistency check.

The increase of $N_{corr}$ during aging can be rather well captured by extrapolating the $N_{corr}(T)$ variation obtained from the growth of the
hump of $\vert \chi_3^{(3)} \vert $ measured at equilibrium above $T_g$ and by translating the $\tau_{\alpha}(t_a)$ in terms of a fictive
temperature $T_{fict}(t_a)$ which decreases during aging, finally reaching $T_w$ when $t_a \gg \tau_{\alpha,eq}$. This yields the continuous line in
Fig. \ref{fig7}, which fairly well captures the data drawn from the aging of $\chi_3^{(3)}$. Because this extrapolation roughly agrees with the aging data,
one can estimate that the quench from $T_g+8$~K to $T_w=T_g-8$~K corresponds to a doubling of $N_{corr,eq}$. The approximately $10\%$ increase reported in Fig. \ref{fig7}
is thus the long time \textit{tail} of this increase, while the first $90\%$ increase cannot be measured because it takes place during the quench.

Beyond the qualitative result that $N_{corr}$ increases during aging, these $\chi_{3}^{(3)}(t_a)$ data can be used to test quantitatively some theories about
the emergence of the glassy state. By gathering, in the inset of Fig. \ref{fig7}, the equilibrium data -symbols lying in the $[1;1.3]$ interval of the
 horizontal axis- and the aging data translated in terms of $T_{fict}(t_a)$ -symbols lying in the $[2;2.3]$ interval-, one extends considerably
the experimental temperature interval, which puts strong constraints onto theories. Summarizing two different predictions
by $\ln(\tau_{\alpha}/\tau_0) = Y N_{corr}^{\psi/3}/(k_BT)$ with $Y\sim T; \psi=3/2$ for Random First
Order Transition theory (RFOT) \cite{RFOT} while $Y\sim 1; \psi = 1$ for the numerical
 approach of Ref. \cite{Cam09}, Fig. \ref{fig7} is designed to test these two predictions -see Ref. \cite{Bru12} for details-:
it shows that both of them are consistent with experiments -contrary to  another prediction relying onto a critical
relation $\tau_{\alpha} \propto N_{corr}^z$, yielding an unrealistic large value of $z \sim 20$ to account for the experiments-.

	\subsection{Strong similarities between third and first cubic susceptibilities} \label{part2-3}

We now come back to equilibrium measurements -i.e. above $T_g$- and compare the behavior of the third-harmonic cubic susceptibility $\chi_{3}^{(3)}$ as well as the first-harmonic cubic susceptibilities $\chi_{3}^{(1)}$ and $\chi_{2;1}^{(1)}$ introduced in Eq. (\ref{eq7}). We remind that $\chi_{2;1}^{(1)}$ corresponds to the case where a static field $E_{st}$ is superimposed to the ac field $E\cos(\omega t)$.

\begin{figure}[t]
\includegraphics[width = 8cm]{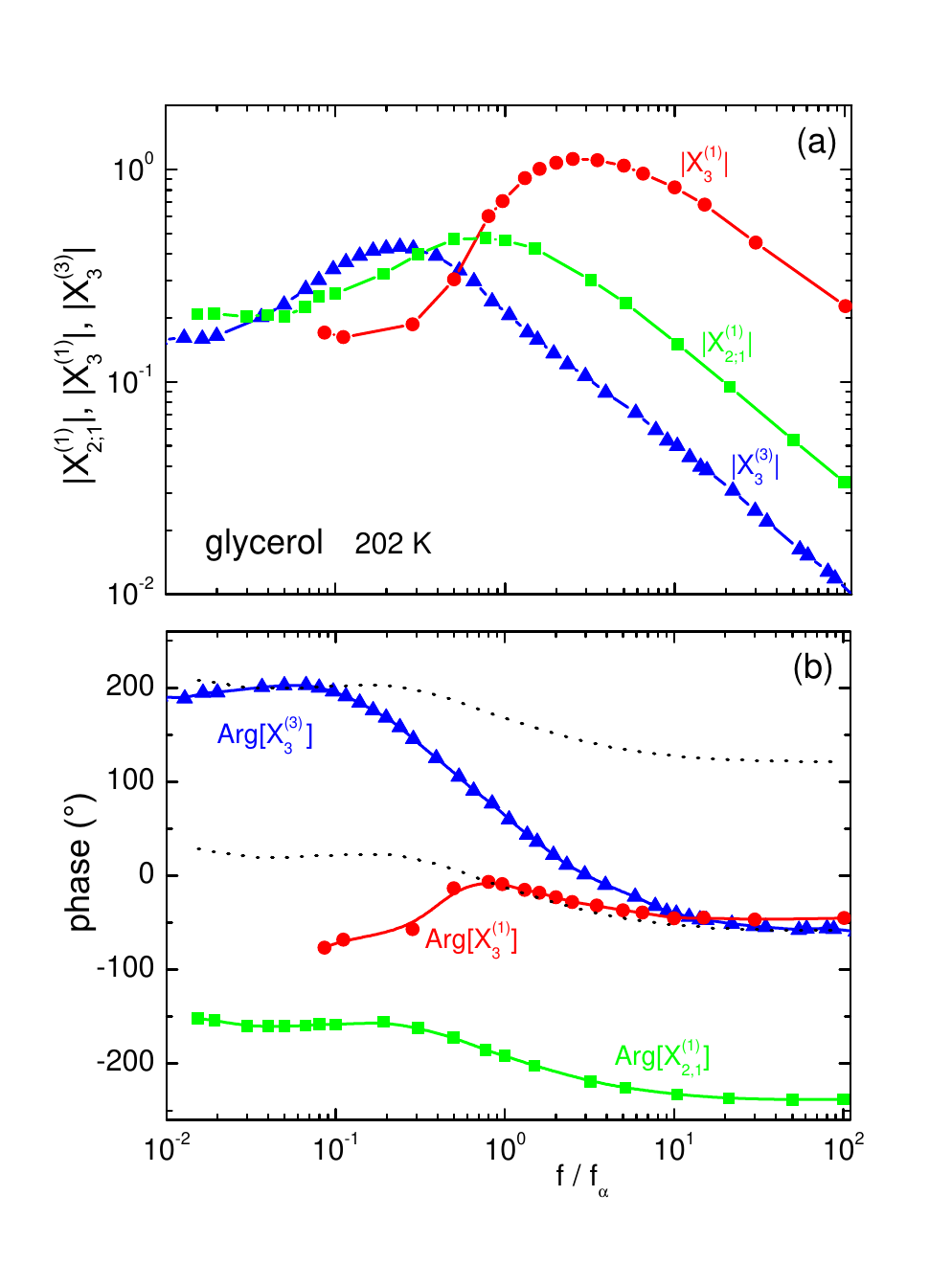}
\caption{From \cite{Gad17}: For glycerol and $f_{\alpha} \simeq 2$~Hz, modulus -top panel- and phase -bottom panel- of the three cubic susceptibilities defined in Eqs. (\ref{eq5}) and (\ref{eq7}). The salient features of the three cubic susceptibilities are similar, which strongly suggests a common physical origin -see text-. Dotted lines are $Arg[X_{2;1}^{(1)}]+\pi$ or $+2 \pi$ and support Eqs. (\ref{eq9}) and (\ref{eq10}).}
\label{fig8}
\end{figure}

\begin{figure}[h]
\includegraphics[width = 8cm]{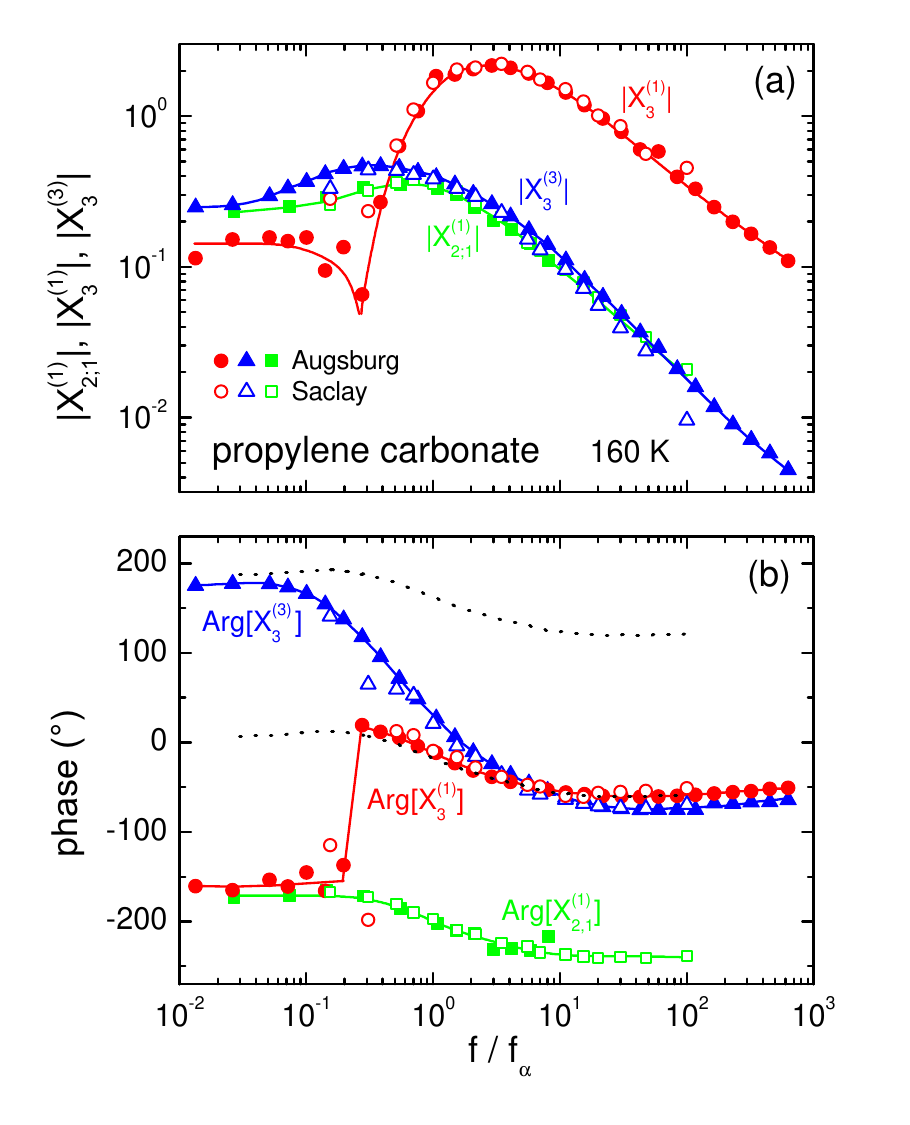}
\caption{From \cite{Gad17}. Same representation as in Fig. \ref{fig8} but for propylene carbonate.}
\label{fig9}
\end{figure}

Figs. \ref{fig8} and \ref{fig9} show the modulus and the phases of the three cubic susceptibilities for glycerol and for propylene carbonate.
\begin{enumerate}

	\item For the modulus: At a fixed temperature, the main features of the frequency dependence of $\vert \chi_{3}^{(1)} \vert$ and of
$\vert \chi_{2;1}^{(1)} \vert$ are the same as those of $\vert \chi_3^{(3)} \vert$: when increasing the frequency, one first observes a low frequency plateau, followed by a hump in the vicinity of $f_{\alpha}$ and then by a power law decrease $\sim f^{-\beta_3}$. The most important differences between the three cubic susceptibilities are the precise location of the hump and the absolute value of the height of the hump. As for the temperature dependence one recovers for  $\vert \chi_{3}^{(1)} \vert$ and for
$\vert \chi_{2;1}^{(1)} \vert$ what we have already seen for $\vert \chi_3^{(3)} \vert$: once put into their dimensionless forms $X_3$ the three cubic susceptibilities do not depend on $T$ in the plateau region, at variance with the region of the hump where they increase upon cooling typically as  $E_{act}(T) \equiv \partial \ln \tau_{\alpha} / \partial (1/T)$ which in this $T$ range is very close to $T \chi_T \equiv \vert \partial \ln \tau_{\alpha} / \partial \ln T \vert$ \cite{Cra10,Bru11,Bau13,Lho14,Cas15,Gad17}.

	\item The phases of the three cubic susceptibilities basically do not depend explicitly on temperature, but only on $u=f/f_{\alpha}$, through a master curve that depends only on the precise cubic susceptibility under consideration. These master curves have the same qualitative shape as a function of $u$ in both glycerol and propylene carbonate. We note that the phases of the three cubic susceptibilities are related to each other. In the plateau region all the phases are equal, which is expected because at low frequency the systems responds adiabatically to the field. At higher frequencies, we note that for both glycerol and propylene carbonate (expressing the phases in radians):

\begin{eqnarray}
\mathrm{Arg} \left[ X_{3}^{(1)} \right] & \approx & \mathrm{Arg} \left[ X_{2,1}^{(1)} \right]+\pi \quad \mathrm{for} f/f_{\alpha} \ge 0.5;  \label{eq9} \\
\mathrm{Arg} \left[ X_{3}^{(1)} \right] & \approx & \mathrm{Arg} \left[ X_{3}^{(3)} \right] \qquad \mathrm{for} f/f_{\alpha} \ge 5  \label{eq10}
\end{eqnarray}
which are quite non trivial relations.

\item In the phase of $\chi_3^{(1)}$ of propylene carbonate (Fig. \ref{fig9}), a jump of $\pi$ is  observed which is accompanied by the indication of a spikelike minimum in the modulus -see \cite{Gad17} for more details-. A similar jump may also be present in glycerol (Fig. \ref{fig8}). This jump in the phase happens at the crossover between the $T$-independent ``plateau''  and the strongly $T$-dependent hump. More precisely in the ``plateau'' region one observes a reduction of the real part of the dielectric constant $\chi_{{1}}'$, while around the hump $\chi_{{1}}'$ is enhanced. At the frequency of the jump, both effects compensate and this coincides with a very low value of the imaginary part of $X_{3}^{(1)}$.

 \end{enumerate}

	\subsection{Frequency and temperature dependence of fifth harmonic  susceptibility} \label{part2-4}

In this section, we first explain why measuring $\chi_5^{(5)}$ is interesting for a better understanding of the glass transition. We then see the characteristic features of $\chi_5^{(5)}$ as a function of frequency and temperature.

	\subsubsection{Interest in the fifth-order susceptibility} \label{part2-4-1}
	
In the previous sections, we have seen  that the increase of the hump
of $\vert X_3 \vert$ upon cooling has been interpreted as reflecting that of
the correlation volume $N_{corr} a^3$. However in practice,
this increase of $N_{corr}$ remains modest -typically it is an
increase by a factor $1.5$- in the range $0.01$~Hz$ \le f_{\alpha} \le10$~kHz
where the experiments are typically performed. Physically
this may be interpreted by the fact that an increase of $N_{corr}$ changes
the activation energy, yielding an exponentially large increase of
the relaxation time $\tau_{\alpha}$. Now if one demands, as in
standard critical phenomena, to see at least a factor of
$10$ of increase of $\vert X_3 \vert$ to be able to conclude
on criticality, one is lead to astronomical values of
$\tau_{\alpha}$: extrapolating the above result,
e.g., $\vert X_3 \vert \propto \vert \partial \ln \tau_{\alpha} /
\partial \ln T \vert$ and assuming a VFT law for $\tau_{\alpha}$,
one concludes that the experimental characteristic times corresponding to an increase of $\vert X_3 \vert$ by one order of magnitude
is $0.1~\textrm{ms} \le \tau_{\alpha} \le 10^{18}~\textrm{s}$. This means experiments  lasting longer than the age of the universe.

This issue of astronomical time scales can be circumvented by using a less commonly exploited but very general property of phase transitions: close to a critical point all the responses diverge together \cite{Kim00}, since the common cause of all these divergences is the growth of the same correlation length. Showing that all the responses of order $k$ behave as a power law of the first diverging susceptibility is another way of establishing criticality. For glasses, we have seen in Eq. (\ref{eq2})  that, apart from $\chi_1$ which is blind to glassy correlations, all other responses $\chi_{k \ge 3}$ grow as power laws with the amorphous ordering length $\ell$: $\chi_3 \propto (l/a)^{2 d_f -d}$ and $\chi_5 \propto (l/a)^{3 d_f -d}$. Therefore, assuming that the main cause for the singular responses appearing in the system is the development of correlations, there should be a scaling relation between the third and fifth order responses, namely one should observe $\chi_5 \propto \chi_3^{\mu (d_f)}$ where $\mu (d_f) = (3 d_f - d)/(2 d_f - d)$.
	
Measuring $\chi_5$ is of course extremely difficult, because, for the experimentally available electric fields, one has the hierarchy $\vert \chi_1 \vert E \gg \vert \chi_3\vert E^3 \gg \vert \chi_5\vert E^5$. However this was done in Ref. \cite{Alb16} and we shall now briefly review the corresponding results.

\begin{figure}[t]
\includegraphics[width = 8cm]{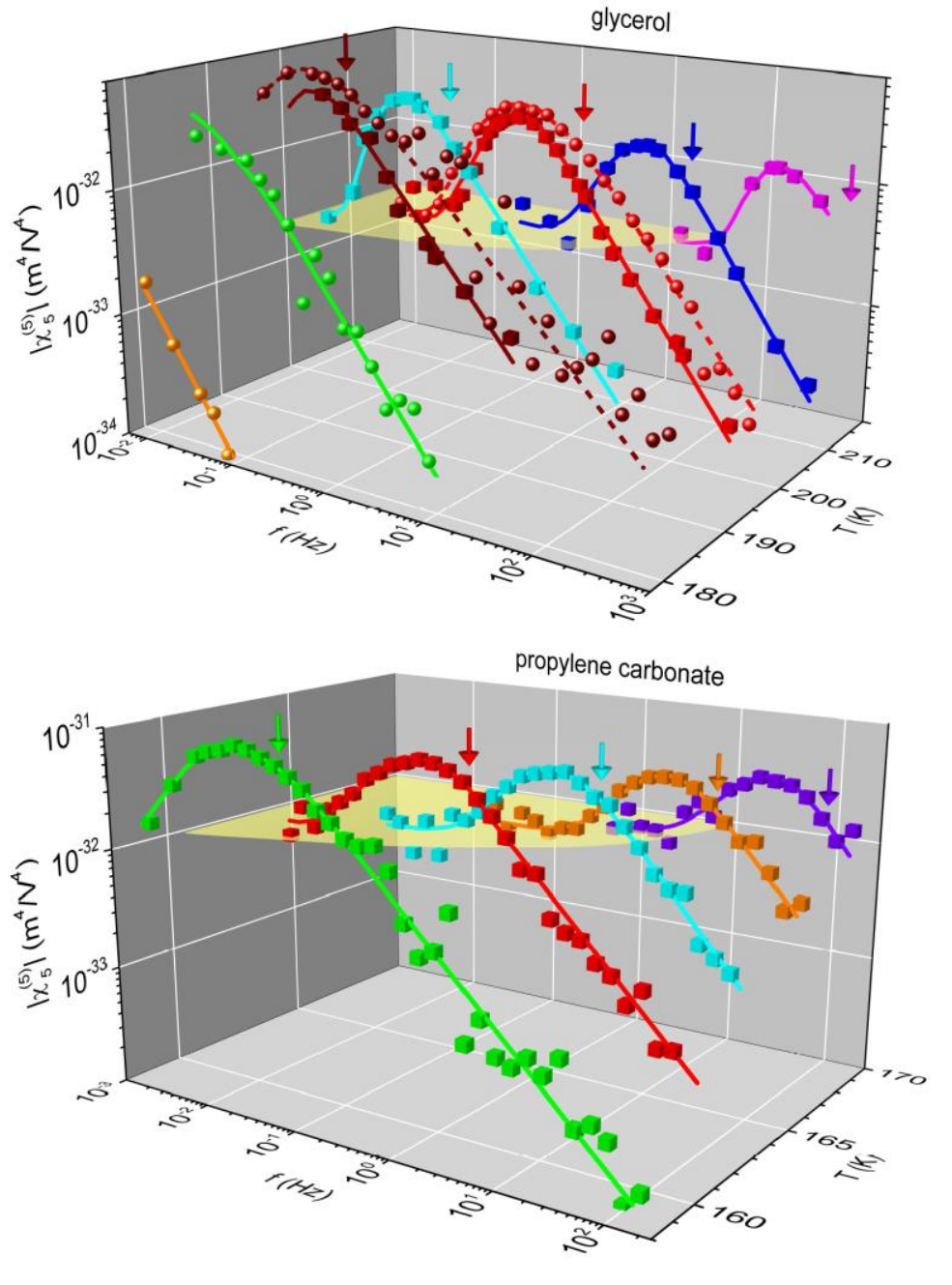}
\caption{From Ref. \cite{Alb16}. Measured values of $\vert \chi_5^{(5)} \vert$ for glycerol - upper panel- and propylene carbonate - lower panel- (the spheres and cubes in the upper panel indicate results from two different experimental setups). The hump lies at the same frequency as for $\vert \chi_3^{(3)} \vert$ and has significantly stronger variations in frequency and in temperature, see Figs \ref{fig11} and \ref{fig12}. The arrows indicate the peak positions $f_\alpha$ in the dielectric loss. The yellow-shaded planes indicate the plateau arising in the trivial regime.}
\label{fig10}
\end{figure}

\begin{figure}[t]
\includegraphics[width = 8cm]{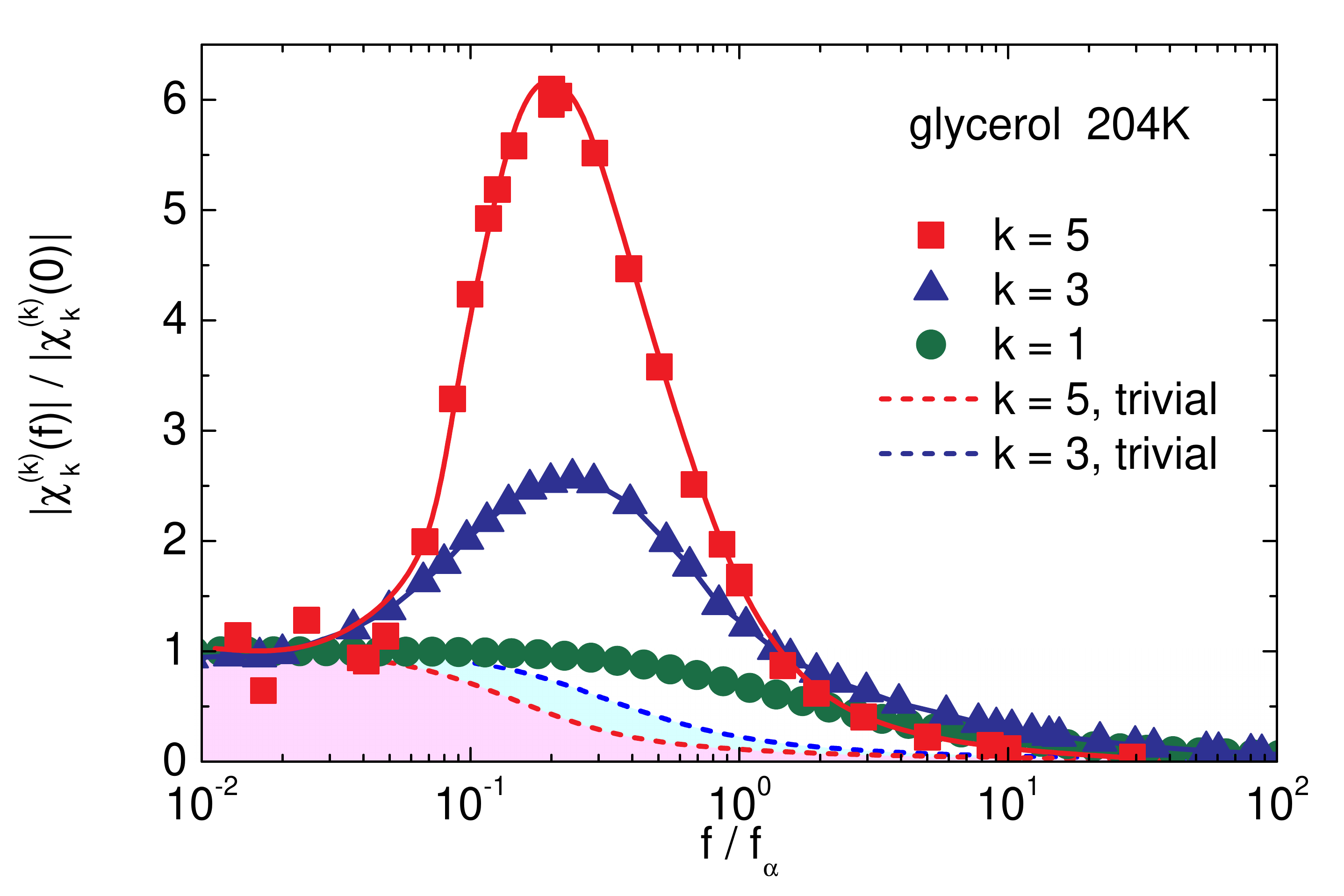}
\caption{From Ref. \cite{Alb16}. For glycerol, comparison of the fifth, third and linear susceptibilities -the latter is noted $\vert \chi_1^{(1)} \vert$-. The hump for $\vert \chi_5^{(5)} \vert$ is much stronger than that of $\vert \chi_3^{(3)} \vert$. The dashed lines are the trivial contribution -see \cite{Alb16} for details-.}
\label{fig11}
\end{figure}

\begin{figure}[h]
\includegraphics[width = 8cm]{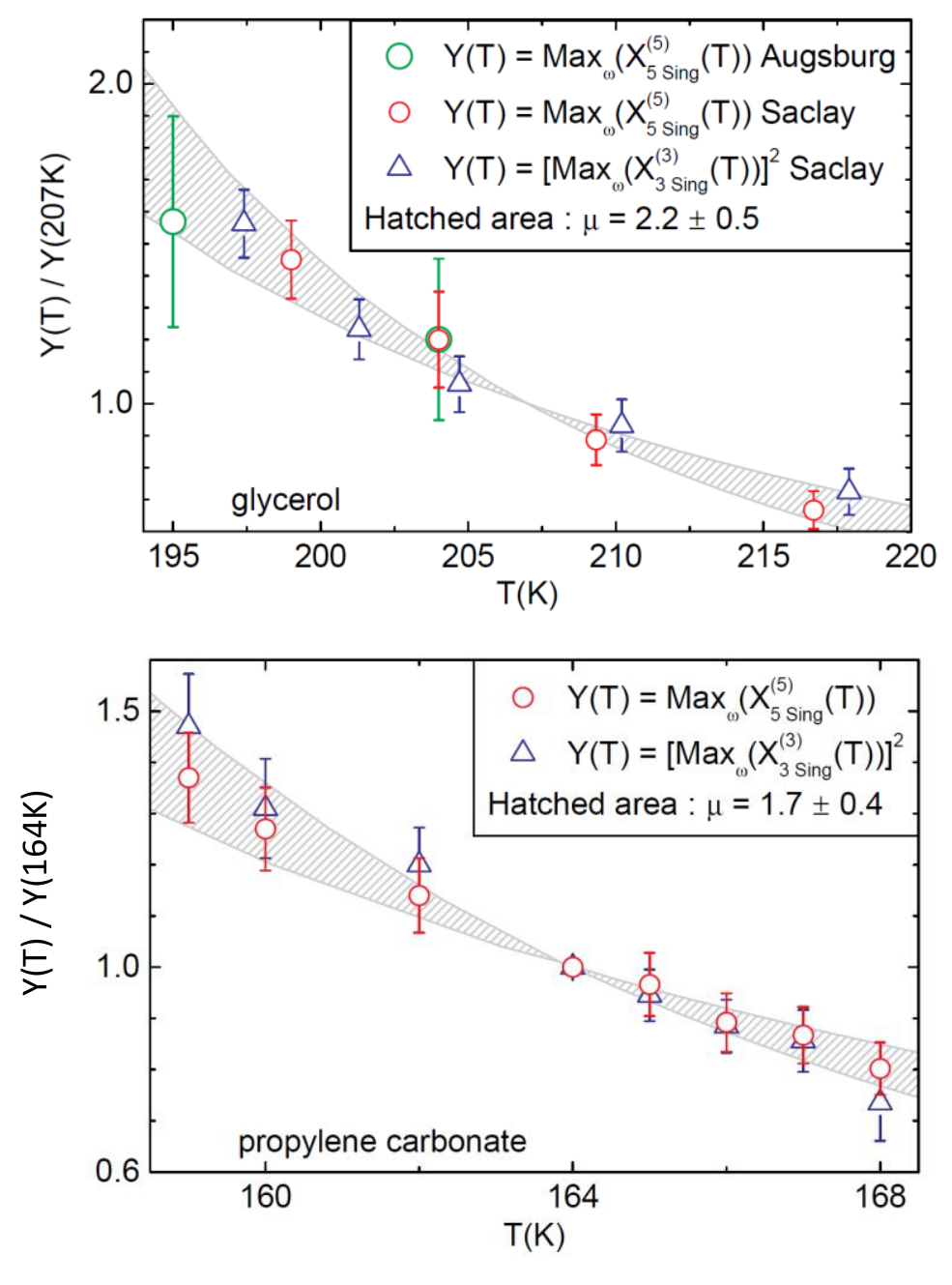}
\caption{From Ref. \cite{Alb16} Temperature evolution of the singular parts of fifth and third order responses. All quantities are normalized at a given temperature -namely $207$~K for glycerol, upper panel; and $164$~K for propylene carbonate, bottom panel-. This allows to determine the exponent $\mu$ relating $\vert X_5 \vert$ and $\vert X_3 \vert^\mu$ and to conclude that the amorphously ordering domains are compact -see text-. The hatched areas represent the uncertainty on $\mu$.}
\label{fig12}
\end{figure}

	\subsubsection{Characteristic features of the fifth order susceptibility}\label{part2-4-2}
	
	The modulus $\vert \chi_5^{(5)}\vert$ of glycerol and propylene carbonate \cite{Alb16} can be seen in Fig. \ref{fig10}  as a function of frequency and temperature. Similarly to what has been seen in section \ref{part2-2} on $\vert \chi_3^{(3)}\vert$, the frequency dependence can be separated in two domains (see also Fig. \ref{fig11}):
	\begin{enumerate}
	\item For very low reduced frequencies ($f/f_\alpha \leq 0.05$), there is a plateau (indicated by the yellow-shaded planes in Fig. \ref{fig10}) where the reduced response $X_5^{(5)}$ depends neither on frequency nor on temperature. In this plateau, the behavior of the supercooled liquid cannot be qualitatively distinguished from the behavior expected from a high temperature liquid of dipoles, depicted by the "trivial" $X_k^{(k)}$ curves represented as dotted lines in Fig. \ref{fig11}.
	\item At higher frequencies, we can observe a hump of $\vert X_5^{(5)}\vert$ that remarkably occurs at the same peak frequency $f_{peak}$ as in $\vert \chi_3^{(3)} \vert$ in both glycerol and propylene carbonate. Again one finds that, for the five temperatures where the peak is studied, $f_{peak} / f_\alpha = c$, where the constant $c$ does not depend on $T$ and weakly changes with the liquid. This peak is \textit{much sharper for} $\vert X_5^{(5)}\vert$ \textit{than for}
 $\vert X_3^{(3)}\vert$: this is clearly evidenced by Fig. \ref{fig11} where the linear, cubic and fifth-order susceptibilities are compared, after normalisation to their low-frequency value. This shows that the anomalous features in the frequency dependence are stronger in $\vert X_5^{(5)} \vert$ than in $\vert X_3^{(3)} \vert$: This may be regarded as a sign of criticality since close to a critical point, the larger the order $k$ of the response, the stronger the anomalous features of $X_k$.
	\end{enumerate}
	
	A second, and more quantitative indication of incipient criticality is obtained by studying the temperature dependence of $\vert X_5^{(5)} \vert$ and by comparing it with that of $\vert X_3^{(3)} \vert$:
	
	\begin{enumerate}
	\item In the plateau region at $f/f_\alpha \leq 0.05$, the value of $\vert X_{5}^{(5)}\vert$ does not depend on the temperature. This shows that the factor involved in the calculation of the dimensionless $X_{5}^{(5)}$ from $\chi_5^{(5)}$ -see Eq. (\ref{eq8})- is extremely efficient to remove all trivial temperature dependences. As the trivial behavior depends on frequency -see the dashed lines of Fig. \ref{fig11}-, the ``singular'' parts of $X_3$ and of $X_5$ are obtained as follows:
\begin{equation}
X_{3{, sing.}}^{(3)} \equiv X_{3}^{(3)} - X_{3 {, trivial}}^{(3)}, \qquad X_{5 {, sing.}}^{(5)} \equiv X_{5}^{(5)} - X_{5 {, trivial }}^{(5)}
\label{eq11c}
\end{equation}
which correspond in Fig. \ref{fig11} to a complex subtraction between the measured data -symbols- and the trivial behavior -dashed lines.
	
	\item Around the hump, the temperature behavior of $\vert X_{5, sing.}^{(5)} (f_{peak})\vert$ is compared to that of $\vert X_{3, sing.}^{(3)} (f_{peak})\vert^\mu$ where $\mu$ is an exponent that is  determined experimentally by looking for the best overlap of the two series of data in Fig. \ref{fig12} -see \cite{Alb16} for details-. This leads us to values of $\mu = 2.2 \pm 0.5$ in glycerol and $\mu = 1.7 \pm 0.4$ in propylene carbonate. Therefore, within experimental uncertainties,  results for $\vert X_{3}^{(3)}\vert$ and $\vert X_{5}^{(5)}\vert$ would seem to advocate a value of $\mu \approx 2$. With $\mu = (3d_f-d)/(2d_f-d)$ as seen in Eq. (\ref{eq2}) -see also Eq. (\ref{eq11b}) below-, this corresponds to a fractal dimensions of $d_f \approx 3$.
	\end{enumerate}

\section{Testing Bouchaud-Biroli's predictions as well as the general theories of the glass transition} \label{part3}
Having shown the experimental data for the nonlinear responses, we now move to the interpretation part and start with Bouchaud-Biroli's approach (BB), which is the most general one. The more specific and/or phenomenological approaches of nonlinear responses will be detailed in Section \ref{part4}.

	\subsection{Bouchaud-Biroli's predictions} \label{part3-1}
	
		\subsubsection{General considerations about $\chi_{2k+1}$}
		
		To illustrate the general relations existing between the susceptibility $\chi_{2k+1}$ and the correlation function of order $2k+2$ -with $k \ge 0$- in a system at thermal equilibrium, let us consider a sample, submitted to a constant and uniform magnetic field $h$,   containing $N$ spins with an Hamiltonian $\cal H$ that depends on the spin configuration ``c''. The elementary relations of statistical physics yield the magnetisation $M \equiv \sum_i <S_i>/(Na^3)$ where $a^3$ is the elementary volume and where the thermal average $<S_i>$ is obtained with the help of the partition function $Z = \sum_{c} \exp{(-\beta {\cal{H}}+\beta h \sum_k S_k)}$ by writing $<S_i> = \sum_c S_i \exp{(-\beta {\cal{H}}+\beta h \sum_k S_k)}/Z$ with $\beta = 1/(k_B T)$. The linear response $\chi_1 \equiv (\partial M / \partial h)_{h=0}$ is readily obtained:

\begin{equation}
N a^3 \chi_1 = \frac{1}{\beta Z} \left(\frac{\partial^2 Z}{ \partial h^2}\right)_{h=0} - \frac{1}{\beta} \left(\frac{\partial Z}{Z \partial h}\right)_{h=0}^2 = \beta \left( \sum_{i1 ; i2} <S_{i1} S_{i2} > - (\sum_{i1} <S_{i1}> )^2 \right)
\label{eq11a}
\end{equation}
which shows that the linear response is related to the connected two-point correlation function. Repeating the argument for higher-order responses -e.g. $\chi_3 \propto (\partial^3 M / \partial h^3)_{h=0}$-, one obtains that $\chi_{2k+1}$ is connected to the $(2k+2)$ points correlation function -e.g., $\chi_3$ is connected to a sum combining $<S_{i1} S_{i2} S_{i3} S_{i4}>$, $<S_{i1} S_{i2} S_{i3}> <S_{i4}>$, $<S_{i1} S_{i2}><S_{i3} S_{i4}>$ etc...-.

		\subsubsection{The spin glass case}
		
		Spin glasses are characterized by the fact that there is frozen disorder, i.e. the set of the interaction constants $\lbrace J_{i;j} \rbrace$ between two given spins $S_i$ and $S_j$ is fixed once and for all, and has a random sign -half of the pairs of spins are coupled ferromagnetically, the other half antiferromagnetically-. Despite the fact that the system is neither a ferromagnet, nor an antiferromagnet, upon cooling it freezes, below a critical temperature $T_{SG}$, into a solid -long range ordered- state called a spin glass state. This amorphous ordering is not detected  by $\chi_1$ which does not diverge at $T_{SG}$: this is because the various terms of $\sum_{i1 ; i2} <S_{i1} S_{i2} >$ cancel since half of them are positive and the other half are negative. By contrast the cubic susceptibility $\chi_3$ contains a term $\sum_{i1 ; i2} <S_{i1} S_{i2} >^2$ which does diverge since all its components are strictly positive: this comes from the fact that the influence $<S_{i1} S_{i2}>$ of the polarization of spin $S_{i1}$ on spin $S_{i2}$ may be either positive or negative, but it has the same sign as the reverse influence $<S_{i2} S_{i1}>$ of spin $S_{i2}$ on spin $S_{i1}$. This is why the amorphous ordering is directly elicited by the divergence of the static value of $\chi_3$ when decreasing $T$ towards $T_{SG}$, as already illustrated in Fig. \ref{fig1}-(A). By adding a standard scaling  assumption close to $T_{SG}$ one can account for the behavior of $\chi_3$ at finite frequencies, i.e. one easily explains that $\chi_{3}$ is frequency independent for $\omega \tau_{\alpha} \le 1$, and smoothly tends to zero at higher frequencies. Finally, similar scaling arguments about correlation functions easily explain the fact that the stronger $k\ge 1$ the more violent the divergence of $\chi_{2k+1}$ in spin glasses, as observed experimentally by Levy \textit{et al} \cite{Lev88}. 		
		
		\subsubsection{The glass forming liquids case}		
The case of glass forming liquids is of course different from that of spin glasses for some obvious reasons (e.g. molecules have both translational and rotational degrees of freedom). As it has been well established that rotational and translational degrees of freedom are well coupled in most of liquids, it is tempting to attempt a mapping between spin glasses and glass forming liquids by replacing the spins $S_i$ by the local fluctuations of density $\delta \rho_i$ or by the dielectric polarisation $p_i$. As far as nonlinear responses are concerned, this mapping requires a grain of salt because
\textit{(a)} there is no frozen-in disorder in glass forming liquids, and \textit{(b)} there is a nonzero value of the molecular configurational entropy $S_c$ around $T_g$.

The main physical idea of BB's work \cite{Bou05} is that these difficulties have an effect which is important at low frequencies and negligible at high enough frequencies:

\begin{enumerate}
\item  Provided $f \ge f_\alpha$, i.e. for processes faster than the relaxation time, one cannot distinguish between a truly frozen glass and a still flowing liquid. If some amorphous order is present in the glass forming system, then non-trivial spatial correlations should be present and lead to anomalously high values of non-linear susceptibilities: this holds for very general reasons -e.g., the Langevin equation for continuous spins which is used in Ref. \cite{Bou05} needs not to specify the detailed Hamiltonian of the system- and comes from an analysis of the most diverging term in the four terms contributing to $\chi_3(\omega)$. If the amorphous correlations extend far enough to be in the scaling regime, one can neglect the subleading terms and one predicts that the nonlinear susceptibilities are dominated by the glassy correlations and given by \cite{Bou05,Alb16}:
\begin{equation}
X_{2k+1}^{{glass}}(f,T) = [N_{{corr}}(T)]^{\alpha_k} \times {\cal H}_k\left(\frac{f}{f_\alpha}\right) \ \hbox{\ with\ }\
\alpha_k = (k+1)- d/d_f
\label{eq11b}
\end{equation}
where the scaling functions ${\cal H}_k$ do not explicitly depend on temperature, but depend on the kind of susceptibility that is considered, i.e. $X_3^{(1)}$, $X_3^{(3)}$
or $X_{2,1}^{(1)}$ in the third order case $k=1$. We emphasize that in Ref. \cite{Bou05} the amorphously ordered domains were assumed to be compact, i.e. $d_f=d$, yielding $\alpha_1 = 1$ i.e. $X_3 \propto N_{corr}$. The possibility of having a fractal dimension $d_f$ lower than the spatial dimension $d$ was considered in Ref. \cite{Alb16} where the fifth order response was studied. As already shown in Section \ref{part2-4-2}, the experimental results were consistent with $d_f=d$, i.e. $X_5 \propto N_{corr}^2$.

\item In the low frequency regime $f \ll  f_{\alpha}$, relaxation has happened everywhere in the system,
destroying amorphous order \cite{note2} and the associated anomalous response to the external field and ${\cal H}_k(0)=0$. In other words, in this very low frequency regime, every molecule behaves independently of others and $X_{2k+1}$ is dominated by the ``trivial'' response of effectively independent molecules.

\end{enumerate}

Due to the definition adopted in Eq. (\ref{eq8}), the trivial contribution to $X_{2k+1}$ should not depend on temperature (or very weakly) . Hence, provided $N_{{corr}}$ increases upon cooling, there will be a regime where the glassy contribution $X_{2k+1}^{{glass}}$ should exceed the trivial contribution, leading to hump-shaped non-linear susceptibilities, peaking at $f_{{{peak}}} \sim f_\alpha$, where the scaling function ${\cal H}_k$ reaches its maximum.

\subsection{Experimentally testing BB's predictions} \label{part3-2}
We now briefly recall why all the experimental features reported in section \ref{part2} are well accounted for by BB's prediction:

\begin{enumerate}

\item The modulus of both the third order susceptibilities $\vert \chi_3^{(3)} \vert, \vert \chi_3^{(1)} \vert, \vert \chi_{2;1}^{(1)} \vert$ and of $\vert \chi_5^{(5)} \vert$ have a humped shape in frequency, contrary to $\vert \chi_{1} \vert$.

\item Due to the fact that ${\cal H}_k$ does not depend explicitly on $T$, the value of $f_{{peak}}/f_{\alpha}$ should not depend on temperature, consistent with the experimental behavior.

\item Because of the dominant role played by the glassy response for $f \ge f_{{peak}}$, the $T$-dependence of $\vert X_{2k+1} \vert$ will be much stronger above $f_{{peak}}$ than in the trivial low-frequency region.

\item Finally, because non-linear susceptibilities are expressed in terms of scaling functions,
it is natural that the behavior of their modulii and phases are quantitatively related especially at high frequency
where the "trivial" contribution can be neglected, consistent with Eqs.  (\ref{eq9})-(\ref{eq10}) --see below for a more quantitative argument in the context of the so-called ``Toy model''-- \cite{note9}.

\end{enumerate}

Having shown that BB's prediction is consistent with experiments, the temperature variation of $N_{corr}$ can be drawn from the increase of the hump of $X_3$ upon cooling. It has been found \cite{Cra10,Bru11,Bau13,Lho14,Cas15} that the temperature dependence of $N_{{corr}}$ inferred from the height of the humps of the three $X_3$'s are compatible with one another, and closely related to the temperature dependence of $T \chi_T$, which was proposed in Refs. \cite{Ber05,Dal07} as a simplified estimator of $N_{{corr}}$ in supercooled liquids. The convergence of these different estimates, that rely on  general, model-free theoretical arguments, is a strong hint that the underlying physical phenomenon is indeed the growth of collective effects in glassy systems -- a conclusion that will be reinforced by analyzing other approaches in Section \ref{part4}.

Let us again emphasize that the BB prediction relies on a scaling argument, where the correlation length $\ell$ of amorphously ordered domains is (much) larger than
the molecular size $a$. This naturally explains the similarities of the cubic responses in microscopically very different liquids such as glycerol and propylene carbonate, as well as
many other liquids \cite{Bau13,Cas15}. Indeed the microscopic differences are likely to be wiped out for large $\ell \propto N_{corr}^{1/d_f}$, much like in usual phase transitions.

\subsection{Static vs dynamic length scale? $\chi_3$ and $\chi_5$ as tests of the theories of the glass transition.} \label{part3-3}

We now shortly discuss whether $N_{corr}$, as extracted from the hump of $\vert X_3 \vert$, must be regarded as a purely dynamical correlation volume, or as a static correlation volume. This ambiguity arises because theorems relating (in a strict sense) nonlinear responses to high-order correlation functions only exist in the static case, and that supplementary arguments are needed to interpret the humped shape of $X_3$ (and of $X_5$) observed experimentally. In the original BB's work \cite{Bou05} it was clearly stated that $N_{corr}$ was a dynamical correlation volume since it was related to a four point time dependent correlation function. This question was revisited in Ref. \cite{Alb16} where it was argued that the experimental results could be accounted for  only when assuming that $N_{{corr}}$ is driven by static correlations. This statement comes from an inspection of the various theories of the glass transition \cite{Alb16}: as we now briefly explain, only the theories where the underlying static correlation volume is driving the dynamical correlation volume are consistent with the observed features of nonlinear responses.

As a first example, the case of the family of kinetically constrained models (KCMs) \cite{KCM} is especially interesting since dynamical correlations, revealed by, e.g., four-point correlation functions, exist even in the absence of a static correlation length. However in the KCM family, one does not expect any humped shape for nonlinear responses \cite{Alb16}. This is not the case for theories (such as RFOT \cite{RFOT} or Frustration theories \cite{Tar05}) where a non-trivial thermodynamic critical point drives the glass transition: in this case the incipient amorphous order allows to account \cite{Alb16} for the observed features of $X_3$ and $X_5$. This is why it was argued in \cite{Gad17,Alb16} that, in order for $X_3$ and $X_5$ to grow, some incipient amorphous order is needed, and that dynamical correlations in strongly supercooled liquids are driven by static (``point-to-set'') correlations \cite{note12} --this statement will be reinforced in section \ref{part4-2}.

\section{More specific models for harmonic susceptibilities} \label{part4}
We now review the various other approaches that have been elaborated for the nonlinear responses of glass forming liquids. We shall see that most of them -if not all- are consistent with BB's approach since they involve $N_{corr}$ as a key -implicit or explicit- parameter.
	
	\subsection{Toy and pragmatical models} \label{part4-1}
	
The ``Toy model'' has been proposed in Refs. \cite{Lho14,Lad12} as a simple incarnation of the BB mechanism, while the ``Pragmatical model'' is more recent \cite{Buc16a,Buc16b}. Both models start with the same assumptions: (i) each amorphously ordered domain is compact and contains $N_{{corr}}$ molecules,  which yields a dipole moment $\propto \sqrt{N_{{corr}}}$ and leads to an anomalous contribution to the cubic response $X_{3}^{\mathrm{glass}} \propto N_{{corr}}$; (ii) there is a crossover at low frequencies towards a trivial cubic susceptibility
contribution $X_{3}^{{triv}}$ which does not depend on $N_{{corr}}$. More precisely, in the ``Toy model'' each amorphously ordered domain is supposed to live in a simplified energy landscape, namely an asymmetric double-well potential with a dimensionless
asymmetry $\delta$, favoring one well over the other. The most important difference between the Toy and the Pragmatical model comes from the description of the low-frequency crossover, see Refs. \cite{Lad12} and \cite{Buc16b} for more details.

On top of $N_{{corr}}$ and $\delta$, the Toy model uses a third adjustable parameter, namely the frequency $f^*$ below which the trivial contribution becomes dominant. In Ref. \cite{Lad12}, \textit{both the modulus and the phase} of $X_3^{(3)}(\omega, T)$ and of $X_3^{(1)}(\omega, T)$ in glycerol were well fitted by using $f^* \simeq f_\alpha/7$, $\delta = 0.6$ and, for $T=204$~K, $N_{{corr}} = 5$ for $X_3^{(3)}$ and $N_{{corr}} = 15$ for $X_3^{(1)}$. Fig. \ref{fig13} gives an example of the Toy model prediction for $X_3^{(3)}$ in glycerol. Besides, in Ref. \cite{Lho14}, the behavior of $X_{2,1}^{(1)}(\omega, T)$ in glycerol was fitted with the same values of $\delta$ and of $f^*$ but with $N_{{corr}} = 10$ (at a slightly different temperature $T=202$~K). Of course,
the fact that a different value of $N_{{corr}}$ must be used for the three cubic susceptibilities reveals that the Toy model is oversimplified, as expected.
However, keeping in mind that the precise value of $N_{{corr}}$ does not change the behavior of the phases, we note that the
fit of the three experimental phases is achieved \cite{Lho14,Lad12} by using the very same values of $f^*/f_{\alpha}$ and of $\delta$. This means
that Eqs. (\ref{eq9}) and (\ref{eq10}) are well accounted for by the Toy model by choosing two free parameters.
This is a quantitative illustration of how the BB general framework does indeed lead to strong relations between the various non-linear susceptibilities, such as those contained in Eqs. (\ref{eq9}) and (\ref{eq10}).

\begin{figure}[h]
\includegraphics[width = 8cm]{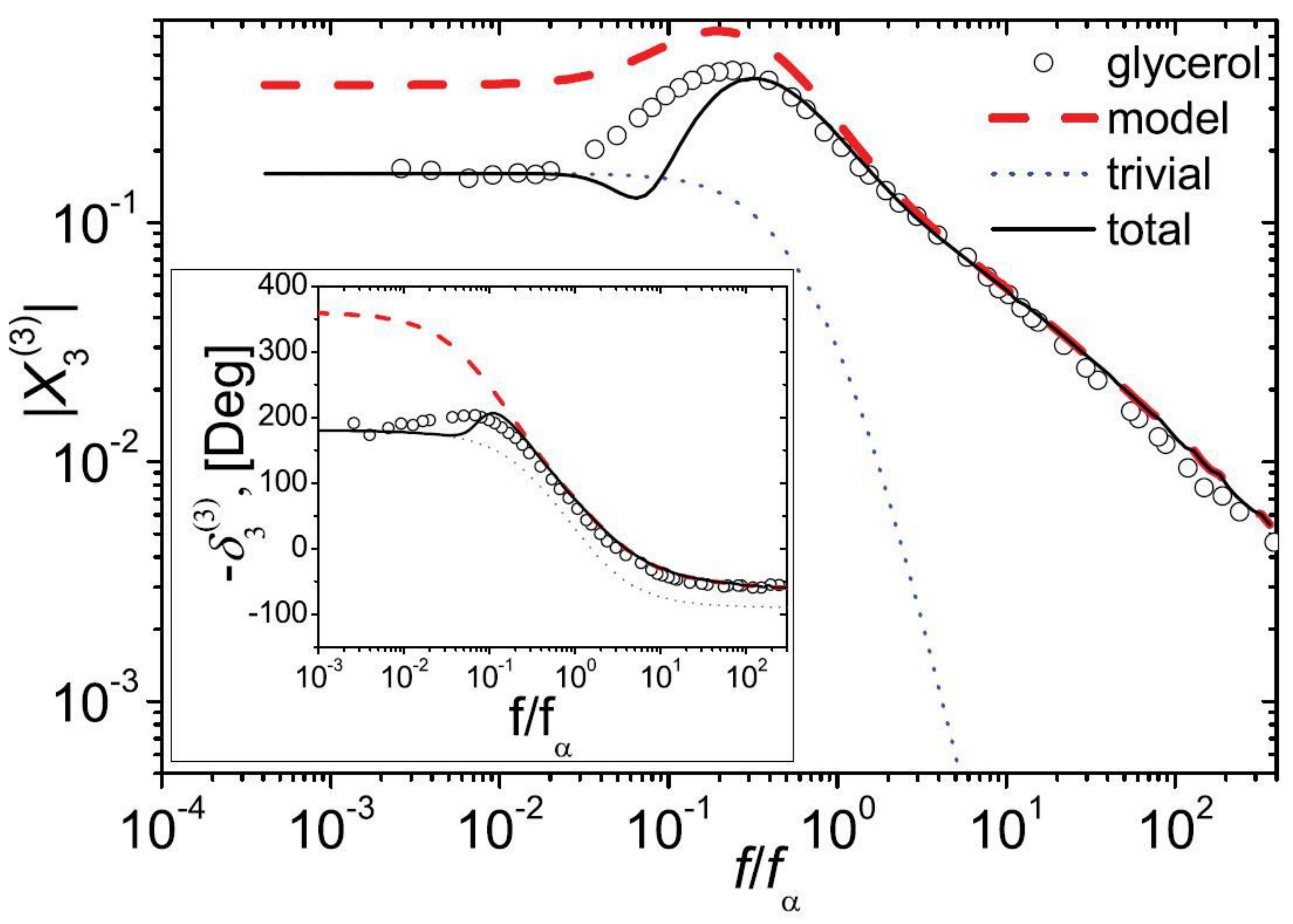}
\caption{From Ref. \cite{Lad12}. Fit of the values of $X_3^{(3)}$ measured in glycerol -symbols- at $204$~K by using the Toy model with $N_{corr} = 5$, $\delta=0.6$ and $f^* \simeq f_\alpha/7$. The prediction of the Toy model is given by the two thick solid lines (main panel for the modulus of $X_3^{(3)}$ and inset for its phase).}
\label{fig13}
\end{figure}

Let us mention briefly the Asymmetric Double Well Potential (ADWP) model \cite{Die12}, which is also about species living in a double well of asymmetry energy $\Delta$, excepted that two key assumptions of the Toy and Pragmatical models are not made: the value of $N_{corr}$ is not introduced, and the crossover to trivial cubic response is not enforced at low frequencies. As a result, the hump for $\vert X_3^{(3)} \vert$ is predicted \cite{Die12,Die17} only when the reduced asymmetry $\delta =\tanh(\Delta/(2k_BT))$ is close to a very specific value, namely $\delta_c = \sqrt{1/3}$, where $X_3$ vanishes at zero frequency due to the compensation of its several terms. However, at the fifth order \cite{Die17} this compensation happens for two values of $\delta$ very different from $\delta_c$: as a result the model cannot predict a hump happening both for the third and for the fifth order in the same parametric regime, contrarily to the experimental results of Ref. \cite{Alb16}. This very recent calculation of fifth order susceptibility \cite{Die17} reinforces the point of view of the Toy and  Pragmatical models, which do predict a hump occurring at the same frequency and temperature due to their two key assumptions ($N_{corr}$ and crossover to trivial nonlinear responses at low frequencies). This can be understood qualitatively: because the Toy model predicts \cite{Lad12} an anomalous contribution $X_{2k+1}^{{glass}} \sim [N_{{corr}}]^k$, provided that $N_{corr}$ is large enough, the magnitude of this contribution is much larger than that of the small trivial contribution $X_{2k+1}^{{triv.}} \sim 1$, and the left side of the peak of $\vert X_{2k+1} \vert$ arises just because the Toy model enforces a crossover from the large anomalous response to the small trivial response at low frequencies $f \ll f_{\alpha}$. As for the right side of the peak, it comes from the fact that $\vert X_{2k+1} \vert \to 0$ when $f \gg f_{\alpha}$ for the simple reason that the supercooled liquid does not respond to the field at very large frequencies.   	
	
	\subsection{Entropic effects} \label{part4-2}

A contribution to nonlinear responses was recently calculated
by Johari in Refs \cite{Joh13,Joh16} in the case where a
static field $E_{st}$ drives the supercooled liquid in
the nonlinear regime. Johari's idea was
positively tested in the corresponding
$\chi_{2;1}^{(1)}$ experiments in
Refs \cite{Sam15,You15,Rie15,Sam16b} -see however Ref. \cite{Sam16} for a case where the agreement is not as good-.
It was then extended to
pure ac experiments -and thus to $\chi_3^{(3)}$-
in Refs. \cite{Ric16a,Ric16b}. The relation between Johari's
idea and $N_{corr}$ was made in Ref. \cite{Gad17}.
		
		\subsubsection{When a static field $E_{st}$ is applied}
		
Let us start with the case
of $\chi_{2;1}^{(1)}$ experiments, i.e. with the case where
a static field $E_{st}$
is superimposed onto an ac field $E \cos(\omega t)$. In this
case, there is a well defined variation of
entropy $\left[ \delta S \right]_{E_{{st}}}$ induced by
$E_{st}$, which, for small $E_{{st}}$ and a fixed $T$, is
given by:
\begin{equation}
\left[\delta S \right]_{E_{{st}}} \approx \frac{1}{2} \epsilon_0 \frac{\partial \Delta \chi_1}{\partial T} E_{{st}}^2 a^3,
\label{eq12}
\end{equation}
where $a^3$ is the molecular volume. Eq. (\ref{eq12}) 
holds generically for any material. However, in the specific
case of supercooled liquids close enough to their glass
transition temperature $T_g$, a special
relation exists between the molecular relaxation
time $\tau_{\alpha}$ and the configurational contribution to
the entropy $S_c$.
This relation, first anticipated by Adam and Gibbs \cite{Ada65},
can be written as :
\begin{equation}
\ln{\frac{\tau_{\alpha}(T)}{\tau_0}}= \frac{{\Delta_0}}{T S_c(T)}
\label{eq13}
\end{equation}
where $\tau_{0}$ is a microscopic time, and $\Delta_0$ is
an effective energy barrier for a molecule. The
temperature dependence of $T S_c(T)$ quite well captures
the temperature variation of $\ln(\tau_{\alpha})$, at least for
a large class of supercooled liquids \cite{Ric98}.

Following Johari \cite{Joh13,Joh16} let us now assume that $\left[\delta S \right]_{E_{{st}}}$ is dominated by the dependence of $S_c$ on field, --see the Appendix of Ref. \cite{Gad17} for a further discussion of this important physical assumption-. Combining Eqs. (\ref{eq12}) and (\ref{eq13}), one finds that a static field $E_{{st}}$ produces a shift of $\ln(\tau_{\alpha}/\tau_0)$ given by:
\begin{equation}
\left[\delta \ln{\tau_{\alpha}} \right]_{E_{{st}}} = -\frac{{\Delta_0}}{T S_c^2} \left[ \delta S \right]_{E_{{st}}}
\label{eq14}
\end{equation}
As shown in Ref. \cite{Gad17} this entropic effect gives
a contribution to $X_{2,1}^{(1)}$, which we call
$J_{2,1}^{(1)}$ after Johari. Introducing
$x=\omega \tau_{\alpha}$, the most general and model-free expression of $J_{2:1}^{(1)}$ reads:
\begin{equation}
J_{2,1}^{(1)} = -\frac{{k_B \Delta_0}}{6S_c^2} \left[\frac{\partial \ln{(\Delta \chi_1)}}{\partial T}\right] \left[ \frac{ \partial \frac{\chi_{{lin}}}{\Delta \chi_1} }{\partial \ln{x}}\right] \propto \frac{1}{S_c^2}
\label{eq15}
\end{equation}
where $\chi_{{lin}}$ is the complex linear susceptibility.

Eq. (\ref{eq15}) deserves three comments:

\begin{enumerate}

 \item $\vert J_{2,1}^{(1)} \vert$ has a
 humped shaped in frequency with a maximum in the region
 of $\omega \tau_{\alpha} \simeq 1$, because
of the frequency dependence of the factor
$\propto \partial \chi_{{lin}}/\partial \ln{x}$ in Eq. (\ref{eq15}).

\item The temperature variation of $J_{2,1}^{(1)}$ is overwhelmingly dominated by that of $S_c^{-1}$
because $S_c \propto (T-T_K)$ -with $T_K$ the Kauzmann temperature-.

\item The smaller $S_c$, the larger must be the size of
the amorphously ordered domains -in the hypothetical limit
where $S_c$ would vanish, the whole sample would be trapped in
 a single amorphously ordered sate and $N_{corr}$ would
 diverge-. In other words, there is a relation between
 $S_c^{-1}$ and $N_{corr}$, which yields \cite{Gad17}:

\begin{equation}
J_{2,1}^{(1)} \propto N_{{corr}}^q,
\label{eq16}
\end{equation}
where it was in shown in Ref. \cite{Gad17} that:

\begin{enumerate}

\item the exponent $q$ lies in the $[2/3;2]$ interval when
one combines the Adam-Gibbs original argument with
general constraints about boundary conditions \cite{Gad17}.

\item the exponent $q$ lies in the $[1/3;3/2]$
interval \cite{Gad17} when one uses the RFOT and plays with its two critical
exponents $\Psi$ and $\theta$. Notably, taking the
``recommended RFOT values'' -$\Psi = \theta = 3/2$ for $d=3$-
gives $q=1$, which precisely corresponds to BB's prediction.
In this case, entropic effects are a physically motivated
picture of BB's mechanism -see \cite{Gad17} for a
refined discussion-.

\end{enumerate}

\end{enumerate}

		\subsubsection{When a pure ac field $E\cos(\omega t)$ is applied}

Motivated by several works \cite{Sam15,You15,Rie15,Sam16b}
showing that Johari's reduction of entropy fairly well
captures the measured $\chi_{2;1}^{(1)}$ in various liquids,
an extension of this idea was proposed in
Refs. \cite{Ric16a,Ric16b} for pure ac experiments, i.e.
for $\chi_{3}^{(3)}$ and $\chi_{3}^{(1)}$. This has given rise
to the phenomenological model elaborated in
Refs. \cite{Ric16a,Ric16b} where the entropy reduction
depends on time, which is nevertheless acceptable in the region $\omega \tau_{\alpha} \le 1$ where the model is used. Fig. \ref{fig14} shows the calculated values for $\vert \chi_3^{(3)} \vert$ at three temperatures for glycerol. The calculation fairly well reproduces the hump of the modulus observed experimentally -the phase has not been calculated-. As very clearly explained in Ref. \cite{Ric16b}, the hump displayed in Fig. \ref{fig14} comes directly from the entropic contribution and not from the two other contributions included in the model (namely the ``trivial'' -or ``saturation''- contribution, and the Box model contribution -see Section \ref{part4-3} below-).

Summarizing this section about entropy effects, we remind the
two main assumptions made by Johari: \textit{(i)} the
field-induced entropy variation mainly goes into
the configurational part of the entropy; \textit{(ii)} its
effects can be calculated by using the Adam-Gibbs relation.
Once combined, these two assumptions give a contribution
to $\chi_{2;1}^{(1)}$ reasonably well in agreement with
the measured values in several
liquids \cite{Sam15,You15,Rie15,Sam16b}. An extension
to $\chi_3^{(3)}$ is even possible, at least in the region
$\omega \tau_{\alpha} \leq 1$ and fairly well accounts for
the measured hump of $\vert \chi_3^{(3)} \vert$ in
glycerol \cite{Ric16a,Ric16b} -a figure similar to
Fig. \ref{fig11} for
$\vert \chi_5^{(5)}(\omega)/\chi_5^{(5)}(0)\vert$ is even
obtained in Ref. \cite{Ric16b}-.
 As shown in Eq. (\ref{eq16}),
this entropy contribution to cubic responses is related
to $N_{corr}$, which is consistent with the general prediction
of BB. Additionally, because $S_c$ is a static quantity,
Eq. (\ref{eq16}) supports the interpretation that the various
cubic susceptibilities $\chi_3$ are related to static
amorphous correlations, as discussed in Section \ref{part3-3}.

\begin{figure}[h]
\includegraphics[width = 8cm]{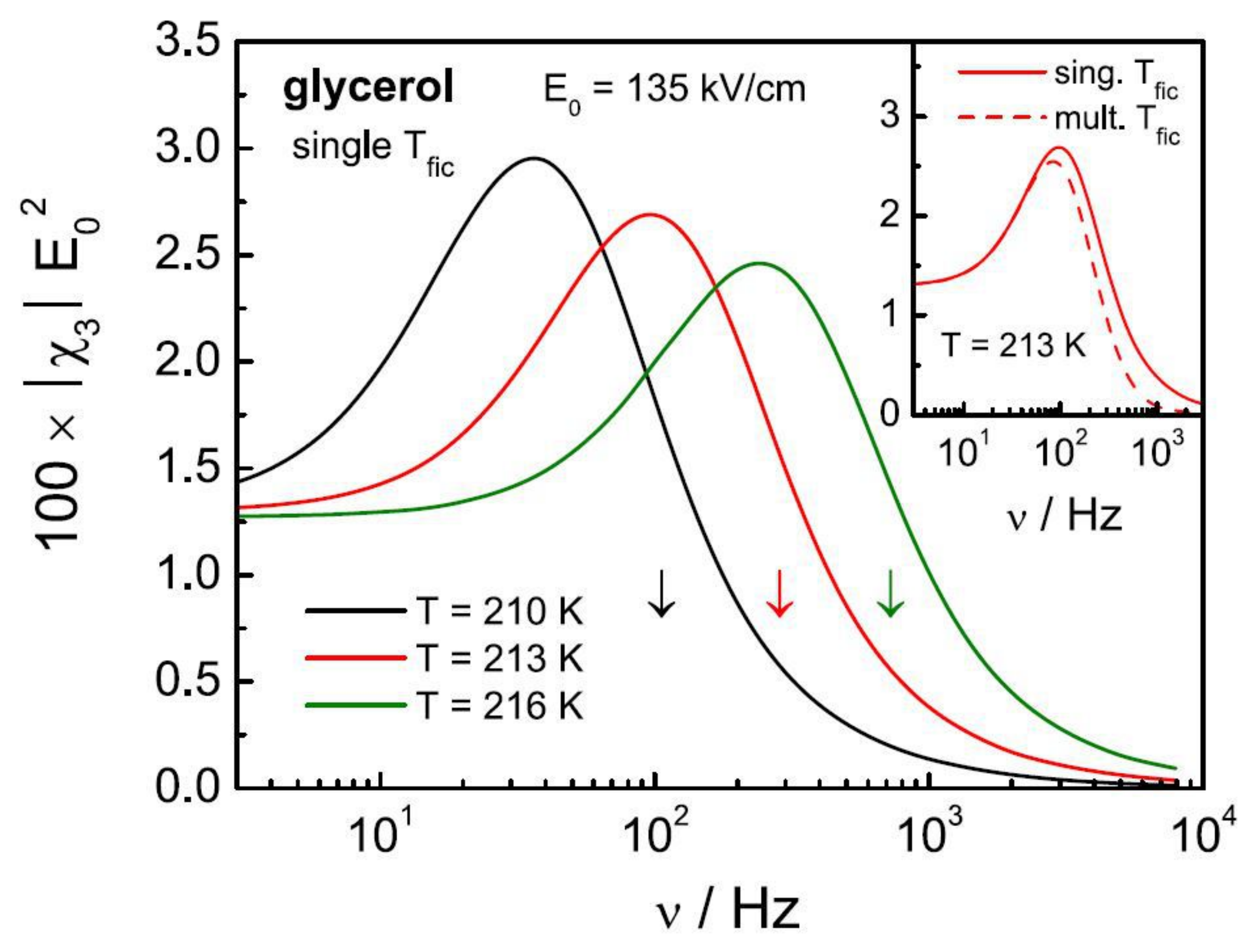}
\caption{From Ref. \cite{Ric16b}. The model elaborated in Refs. \cite{Ric16a,Ric16b} includes three contributions
-entropy reduction, Box model, and trivial-. It predicts for $\vert \chi_3^{(3)} \vert$ the solid lines which account
very well for the measured values in glycerol in frequency and in temperature. The peak of $\vert \chi_3^{(3)} \vert$ arises
because of the entropy reduction effect (noticed ``sing. $T_{fic.}$'') which completely dominates the two other contributions
in the peak region, as shown by the inset.}
\label{fig14}
\end{figure}

	\subsection{Box model} \label{part4-3}
			\subsubsection{Are nonlinear effects related to energy absorption ?} \label{part4-3-1}

The ``Box model'' is historically the first model of
nonlinear response in supercooled liquids, designed to account
for the Nonresonant Hole Burning (NHB) experiments \cite{Sch96}. When these pionneering experiments were carried out, a central question was  whether the dynamics in supercooled liquids is homogeneous or heterogeneous.  In the seminal ref. \cite{Sch96} it was reported that when  applying a strong ac field $E$ of angular frequency $\omega$, the changes in the dielectric spectrum are localised close to $\omega$ and that they last a time of the order of $1/\omega$. These two findings yield a strong qualitative support to the heterogenous character of the dynamics, and the Box model was designed to provide a quantitative description of these results. Accordingly, the Box model assumes that the dielectric response comes from ``domains'' -that will be later called Dynamical Heterogeneities (DH)-, each domain being characterized by its dielectric relaxation time $\tau$ and obeying the Debye dynamics. The distribution of the various $\tau$'s is chosen to recover the measured non Debye spectrum by adding the various linear Debye susceptibilities $\chi_{1,dh} = \Delta \chi_1/(1-i \omega \tau)$ of the various domains. For the nonlinear response, the Box model assumes that it is given by the Debye linear equation in which $\tau(T)$ is replaced by $\tau(T_f)$ where the fictive temperature $T_f= T + \delta T_f$ is governed by the constitutive equation -see e.g. \cite{Ric06,Bru11b}-:

\begin{equation}
c_{dh} \frac{\partial (\delta T_f)}{\partial t} + \kappa \delta T_f = \frac{1}{2} \epsilon_0 \chi''_{1,dh} \omega E^2  
\label{eqBM1}
\end{equation}
with $c_{dh}$ the volumic specific heat of the DH under 
consideration, $\kappa$ the thermal conductance (divided by 
the DH volume $v$) between the DH and 
the phonon bath, $\tau_{therm} = c/\kappa$ the corresponding thermal 
relaxation time. In Eq. (\ref{eqBM1}), only the constant part of the dissipated power has been written, omitting its component at $2\omega$ which is important only for $\chi_3^{(3)}$ -see e.g. \cite{Bru11b}-. From Eq. (\ref{eqBM1}) one easily finds  the stationary value $\delta T_f^\star$ of $\delta T_f$ which reads:
\begin{equation}
\delta T_f^\star = \frac{\tau_{therm}}{\tau} \frac{\epsilon_0 \Delta \chi_1 E^2}{2c_{dh}}\frac{\omega^2 \tau^2}{1+\omega^2 \tau^2}
\label{eqBM1bis}
\end{equation}
As very clearly stated in the seminal 
Ref. \cite{Sch96} because the DH size is smaller than $5$nm, the 
typical value of $\tau_{therm}$ is at most in the nanoseconds range: 
this yields, close to $T_g$, a vanishingly small  value of $\tau_{therm}/\tau$, 
which, because of Eq. (\ref{eqBM1bis}), gives fully negligible values 
for $\delta T_f^\star$. The choice 
of the Box model is to increase $\tau_{therm}$ by orders of magnitude 
by setting $\tau_{therm} = \tau$, expanding onto the intuition that 
this is a way to model the ``energy storage'' in the 
domains. The main justification of this choice is \textit{its efficiency}: 
it allows to account reasonably well for the NHB experiments 
\cite{Sch96} and thus to bring a strong support to the 
heterogeneous character of the dynamics in supercooled liquids. Since 
the seminal ref. \cite{Sch96}, some other works have
 shown \cite{Ric06,Wei07,Wan07,Kha11} that the Box model 
 efficiently accounts for the measured $\chi_{3}^{(1)}(f>f_{\alpha})$ in many glass forming liquids. It was shown
 also \cite{Bru11b} that the Box model is \textit{not} able to
fit quantitatively the measured $X_3^{(3)}$ (even though
some qualitative features are accounted for), and that the
Box model only provides a vanishing contribution to
$X_{2,1}^{(1)}$  -- see \cite{Lho14}.

The key choice $\tau_{therm} = \tau$ made by the Box model has 
two important consequences for cubic susceptibilities:  
it implies \textit{a)} that $\chi_{3}^{(1)}$ mainly comes from the 
energy absorption (since the source term in Eq. (\ref{eqBM1}) is the
 dissipated power) and \textit{b)} that $\chi_{3}^{(1)}$ does 
 not explicitly depend on the 
 volume $v = N_{corr} a^3$ of the DH's (see \cite{Ric06,Bru11b}). 
 However, alternative models of nonlinear 
 responses are now available \cite{Lad12,Buc16b} where, instead 
 of choosing $\tau_{therm}$, one 
 directly resolves the microscopic population
equations, which is a molecular physics approach, 
and not a macroscopic law transferred to microscopics. The 
population equations approach is equivalent to solving the 
relevant multidimensional Fokker-Planck equation describing the 
collective tumbling dynamics of the system at times longer than the 
time between two molecular collisions 
(called $\tau_c$ in Appendix 3). By using this molecular physics 
approach one obtains that $\chi_3^{(1)}$ is governed by $N_{corr}$ and not by energy absorption. For $\chi_3^{(1)}$, writing loosely $P_3^{(1)} \approx \partial P_1 /(\partial \ln \tau) \delta \ln \tau$, one sees that the pivotal quantity is the field induced shift of the 
 relaxation time $\delta \ln \tau$. Comparing the Box model (BM) and, e.g., the Toy model (TM), one gets respectively:
 
 \begin{equation}
 \delta \ln \tau_{BM} \simeq -\frac{1}{2} \chi_T \frac{\epsilon_0 \Delta \chi_1 E^2}{c_{dh}} \quad ; \quad  \delta \ln \tau_{TM} \simeq -\frac{3}{2} \frac{N_{corr}}{T}\frac{\epsilon_0 \Delta \chi_1 E^2}{k_{B}/a^3}
 \label{eqBM2}
 \end{equation}
where we remind our definition $\chi_T =\vert \partial \ln \tau_{\alpha} / \partial T \vert$ and where the limit $\omega \tau \gg 1$, relevant for the $\chi_3^{(1)}(f>f_{\alpha})$  was taken in the Box model, while the simplest case (symmetric double well with a net dipole parallel to the field) was considered for the Toy model. Eq. (\ref{eqBM2}) deserves two comments:

\begin{enumerate}

\item one sees that the two values of $\delta \ln \tau$ are 
 similar provided  $N_{{corr}}$ and $T \chi_T$ are proportional
  -- which is a reasonable assumption as explained  above
  and in Refs. \cite{Bru11,Ber05,Dal07}. Taking reasonable values of this proportionality factor, it was shown 
  in Ref. \cite{Gad17} that $\chi_3^{(1)} (f>f_{\alpha})$ is \textit{the same} in the two models. This sheds a new light on the efficiency of the Box model and on consequence \textit{b)}.
  
\item Let us shortly discuss consequence \textit{a)}. In the Toy model, 
$\delta \ln \tau$ directly expresses 
the field induced modification of the energy of each of the two 
wells modeling a given DH. It comes from the work produced by $E$ onto 
the DH and this is why it involves $N_{corr}$: the larger this number, 
the  larger the work produced by the field because the net dipole of a DH is $\propto \sqrt{N_{corr}}$ and thus increases with $N_{corr}$. It is easy to show
 that the dissipation -i.e. the ``energy absorption''- is \textit{not}  involved 
 in $\delta \ln \tau$ because dissipation depends only on 
 $\chi''_{1}$, which in the Toy model does not depend on $N_{corr}$. 
 In the Toy model, as in the Pragmatical model \cite{Buc16b} and 
 the Diezemann model \cite{Die12}, the heating is neglected because at 
 the scale of a given DH it is vanishingly small as shown above 
 when discussing $\tau_{therm}$. Of course, at the scale of the 
 whole sample, some global heating arises for thick samples and/or 
 high frequencies because the dissipated power has to travel to 
 the electrodes which are the actual heat sinks in dielectric 
 experiments \cite{Bru10}. This purely exogeneous effect can be precisely calculated by solving the heat propagation equation, see e.g. ref.\cite{Bru10} and Appendix 2, and must not be confused with what was discussed in this section.

\end{enumerate}

			\subsubsection{Gathering the three measured cubic susceptibilities} \label{part4-3-2}

As explained above, in Refs. \cite{Ric16a,Ric16b},
the three experimental
cubic susceptibilities have been argued to result from
 a superposition of an entropic contribution and of an 
energy absorption contribution coming from the Box model (plus a
 trivial contribution playing a minor role around the peaks of
 the cubic susceptibilities). More precisely, the hump
 of $\vert X_{2,1}^{(1)} \vert$ and of
 $\vert X_3^{(3)} \vert$ would be mainly due to the
 entropy effect, contrarily to the hump of
 $\vert X_{3}^{(1)} \vert$ which would be due to the Box
 model contribution. As noted in Ref. \cite{Gad17}, this
 means that very different physical mechanisms
 would conspire to give contributions of
 the same order of magnitude, with phases that have
 no reason to match as they do empirically, see
 Eqs. (\ref{eq9}) and (\ref{eq10}): why should $X_3^{(1)}$ and
 $X_3^{(3)}$ have the same phase at high frequencies if
 their physical origin is different?

This is why it was emphasized in Ref. \cite{Gad17}
that there is  no reason for such a similarity if the growth of
$X_3^{(1)}$ and $X_3^{(3)}$ are due to independent
mechanisms. Because entropic effects have been related to the
increase of $N_{{corr}}$ -see
Eqs. (\ref{eq15}) and (\ref{eq16})-, everything becomes instead
very natural if the Box model is recasted in a framework
where $X_3^{(1)}$ is related to the glassy correlation volume.
As evoked above, a first step in this direction was done in
Ref. \cite{Gad17} where it was shown that the Box
model prediction for $X_{3}^{(1)}$ at high frequencies
is identical to the above Toy model prediction,
provided  $N_{{corr}}$ and $T \chi_T$ are proportional.
 In all, it is argued in Ref. \cite{Gad17} that
 the only reasonable way to account for the similarity
  of all three cubic susceptibilities, demonstrated
  experimentally in Figs. \ref{fig8} and \ref{fig9}, is to invoke a
common physical mechanism. As all the other existing
approaches, previously reviewed, relate
cubic responses to the growth of the glassy correlation
volume, reformulating the Box model along the same line seems
to be a necessity.

\section{Conclusions}

We have reviewed in this chapter the salient features reported for
the third and fifth harmonic susceptibilities close to the
glass transition. This is a three decades long story, which has started
in the mid-eighties as a decisive tool to evidence the solid, long
range ordered, nature of the spin glass phase. The question of
whether this notion of ``amorphous order'' was just a
curiosity restricted to the -somehow exotic- case of spin glasses
remained mostly theoretical until  the seminal work
of Bouchaud and Biroli in 2005. This work took a lot from
the spin glass physics, and by taking into account the
necessary modifications relevant for glass forming liquids, it
has  anticipated all the salient features discovered in
the last decade for the three
cubic susceptibilities $X_3$. This is why,
in most of the works, the increase of the hump of $X_3$ upon cooling
has been interpreted as reflecting that of the glassy
correlation volume. Challenging alternative and more specific
 interpretations have been proposed,
but we have seen  that most -if not all- of them can be
recasted into the framework of BB. The avenue opened by BB's
prediction was also used to circumvent the issue of exponentially
long time scales -which are the reason why the nature of the
glass transition is still debated-: this is how the idea of
comparing the anomalous features of $X_3$ and of $X_5$ has arisen.
The experimental findings are finally consistent with the existence of
an underlying thermodynamic critical point, which drives the
formation of amorphously ordered compact domains, the size of
which increases upon cooling. Last we note that this field of
nonlinear responses in supercooled liquids has been inspiring both theoretically \cite{Zan08,Bra10} and  
experimentally, e.g. for colloidal glasses: the very
recent experiments \cite{Sey16}
have shed a new light on the colloidal glass transition and
shown interesting differences with glass forming liquids.

All these progresses open several routes of research. On the
purely theoretical side, any prediction of nonlinear responses in one
of the models belonging to the Kinetically Constrained Model family
will be extremely welcome to go beyond the general arguments given
in Refs. \cite{Alb16}. Moreover, it would be very interesting
to access $\chi_3$ (and $\chi_5$) in molecular liquids at
higher temperatures, closer to the Mode Coupling Transition
temperature $T_{MCT}$, and/or for frequencies close to the fast
$\beta$ process
where more complex, fractal structures with $d_f<d$ may be
anticipated \cite{IMCT,WolynesSchmalian}. This will require a
joined effort of experimentalists -to avoid heating issues- and
of theorists -to elicit the nature of nonlinear responses close
to $T_{MCT}$-. Additionally, one could revisit the vast field of
polymers by monitoring their nonlinear responses, which should shed
new light onto the temperature evolution of the correlations in
these systems. Therefore there is likely much room to deepen
our understanding of the glass transition by carrying out
new experiments about nonlinear susceptibilities.\\

{\bf{ACKNOWLEDGEMENTS}}
We thank C. Alba-Simionesco, Th. Bauer, U. Buchenau, A. Coniglio, G. Johari, K. Ngai,   R. Richert, G. Tarjus, and M. Tarzia for interesting
discussions. The work in Saclay has been supported by the Labex RTRA grant Aricover and by the Institut des Syst\`emes
Complexes ISC-PIF. The work in Augsburg was supported by the Deutsche Forschungsgemeinschaft via Research Unit FOR1394.

\section{Appendix 1: making sure that exogeneous effects are negligible} \label{part5}
We briefly explain how the nonlinear effects reported here
have been shown to be -mainly- free of exogeneous effects:

\begin{enumerate}

	\item The global homogeneous heating of the samples by the dielectric energy dissipated by the application of the strong ac field $E$ was shown to be fully negligible for $X_3^{(3)}$ as long as the inverse of the relaxation time $f_{\alpha}$ is $\le 1$~kHz, see Ref. \cite{Bru10}. Note that these homogeneous heating effects contribute much more to $X_3^{(1)}$: to minimize them, one can either keep $f_{\alpha}$ below $10$~Hz  \cite{Bru11}, and/or severely limit the number $n$ of periods during which the electric field is applied -see, e.g., \cite{Ric06,Sup13}).

\item The contribution of electrostriction was demonstrated
 to be safely negligible in Refs. \cite{Wei07,Bru11}, both by
 using theoretical estimates and
by showing that changing the geometry of spacers does not
affect $X_3^{(3)}$.

\item As for  the small ionic impurities present in most of liquids, we briefly explain that they have a negligible role, except at zero frequency where the ion contribution might explain why the three $X_3$'s are not strictly equal, contrarily to what is expected on general grounds -see, e.g., Figs. \ref{fig8} and \ref{fig9}-. On the one hand  it was shown that the ion heating contribution  is fully negligible in $X_{2,1}^{(1)}$ (see Ref. \cite{Lho14}), on the other hand it is well known that ions affect the linear response $\chi_{{1}}$ at very low frequencies (say $f/f_{\alpha} \le 0.05$): this yields an upturn on the out-of-phase linear response $\chi^{\prime \prime}_{{1}}$, which diverges as $1/\omega$ instead of vanishing as $\omega$ in an ideally pure liquid containing only molecular dipoles.  This may be the reason why most of the $\chi_3$ measurements are reported above  $0.01 f_{\alpha}$: at lower frequencies the nonlinear responses  is likely to be dominated by the ionic  contribution.

\end{enumerate}

\section{Appendix 2: Trivial third and fifth harmonic susceptibilities} \label{part6}
As explained in the main text, in the long time limit -i.e. for $f/f_{\alpha} \ll 1$-, the liquid flow destroys the glassy correlations, making each molecule
 effectively independent of others. This is why we briefly recall what the nonlinear responses of an ideal gas of dipoles are, where each dipole is independent
 of others, and undergoes a Brownian rotational motion -of characteristic time $\tau_D$- due to the underlying thermal reservoir at temperature $T$. The linear
 susceptibility of such an ideal gas of dipoles is given by the Debye susceptibility $\Delta \chi_1/(1-i \omega \tau_D)$, hence the subscript ``Debye'' in the
 Eq. (\ref{eqDebye}) below.
 By using Refs. \cite{Cof76}, and following the definitions given in the main text, as well as Eqs. (\ref{eq5})-(\ref{eq8}) above, one gets for the dimensionless nonlinear responses of such an
 ideal gas, setting for brevity $x=\omega \tau_D$:

\begin{eqnarray}
X_{3 {, Debye}}^{(3)} & = & \left( \frac{-3}{5}\right) \frac{{3-17 x^2 + i x (14-6x^2)}}{{(1+9x^2)(9+4x^2)(1+9x^2)}} \nonumber  \\
X_{5 {, Debye}}^{(5)}& = &  \frac{432(72-2377 x^2 -1979 x^4 + 2990 x^6)}{1680(1+x^2)(4+x^2)(9+4x^2)(1+9x^2)(9+16x^2)(1+25x^2)} \nonumber \\
\ \              &\  & +i \frac{432x(246-737 x^2 -1623 x^4 + 200 x^6)}{560(1+x^2)(4+x^2)(9+4x^2)(1+9x^2)(9+16x^2)(1+25x^2)}
\label{eqDebye}
\end{eqnarray}

In Ref. \cite{Alb16} the trivial response combined the
above $X_{k {,Debye}}^{(k)}$ with a
distribution ${\cal G} (\tau)$ of relaxation times $\tau$
chosen to account for the linear susceptibility of
the supercooled liquid of interest. In Refs \cite{Lad12,Lho14}
a slightly different modelization was used since
${\cal G} (\tau)$ was replaced by the Dirac delta function
$\delta(\tau - \tau_{\alpha})$, i.e. $\tau_D$ was
simply replaced by $\tau_{\alpha}$ for the cubic
trivial susceptibilities.

\section{Appendix 3: Derivation of the Toy model from Langevin Fokker- Planck considerations} \label{part7}
In this section we shall rederive the phenomenological Toy model of Ladieu et al. \cite{Lad12} starting  from the Langevin-Fokker-Planck equation, which is the starting point of Bouchaud and Biroli when they illustrate their general theoretical ideas in the last part of Ref. \cite{Bou05}. We shall idealize the supercooled state of a liquid as follows. At high temperatures, the liquid is made of molecules the interactions between which are completely negligible. On cooling, the molecules arrange themselves in groups, called ``dynamical heterogeneities'' (DH), between which there are no interactions. Inside a typical group, specific intermolecular interactions manifest themselves dynamically, by which we mean that in a time larger than a characteristic time $\tau_{\alpha}$, such interactions lose their coherence and the typical behavior of the liquid is that of an ideal gas. Before and around $\tau_{\alpha}$, these interactions manifest themselves in a frequency range $\omega\approx 1/\tau_{\alpha}$. Thus, stricto sensu, our modelling of this specific process pertains to the behavior of the various dielectric responses of a DH, linear and nonlinear, near this frequency range. This indeed implies that information regarding the ``ideal gas'' phase must be added to fit experimental data. It may be shown on fairly general grounds that either for linear and non-linear responses, such extra information simply superposes onto the specific behavior that has been alluded to above \cite{Dej18}.
Now, we consider that a) a given DH has a given size at temperature $\it{T}$, b) that a DH is made of certain mobile elements that do interact between themselves, c) that there are no interactions between DHs, d) that the dipole moment of a DH is $\mu_{d} = \mu\sqrt{\it{N_{corr}}}$, and e) that all constituents of a DH are subjected to Brownian motion.
\newline
In order to translate the above assumptions in mathematical language, we assign to each constituent of a DH a generalized coordinate $\it{q_i}(\it{t})$, so that each DH is described by a set of generalized coordinates $\textbf{q}$ at temperature $\textit{T}$, viz.
\begin{eqnarray}
\mathbf{q}\left( t \right)=\left\{ {{q}_{1}}\left( t \right),\ldots ,{{q}_{n}}\left( t \right) \right\}
\nonumber
\end{eqnarray}
Inside each DH, each elementary constituent is assumed to interact via a multidimensional interaction potential $V_{int}(\textbf{q})$ that possesses a double-well structure with minima at $\textbf{q}_{\textit{A}}$ and  $\textbf{q}_{\textit{B}}$, and are sensitive both to external stresses and thermal agitation. The equations of motion may be described by overdamped Langevin equations with additive noise, viz.
\begin{eqnarray}
{{\dot{q}}_{i}}=-\frac{1}{\zeta }\frac{\partial {{V}_{T}}}{\partial {{q}_{i}}}\left( \mathbf{q},t \right)+{{\Xi }_{i}}\left( t \right)
\label{eq1A3}
\end{eqnarray}
where $\zeta$ is a generalized friction coefficient, $V_{T}=V_{int}+V_{ext}$, $V_{ext}$ is the potential energy of externally applied forces and the generalized forces ${{\Xi }_{i}}\left( t \right)$ have Gaussian white noise properties, namely
\begin{eqnarray}
\overline{{{\Xi }_{i}}\left( t \right)}=0,
\overline{{{\Xi }_{i}}\left( t \right){{\Xi }_{j}}\left( {{t}'} \right)}=\frac{2kT}{\zeta }{{\delta }_{ij}}\delta \left( t-{t}' \right)
\label{eq2A3}
\end{eqnarray}
Thus, the dynamics of a DH is represented by the stochastic differential equations (\ref{eq1A3}) and (\ref{eq2A3}), which are in effect the starting point of the Bouchaud-Biroli theory, as stated above. A totally equivalent representation of these stochastic dynamics is obtained by writing down the Fokker-Planck equation \cite{Ris89} for the probability density $W\left( \mathbf{q},t \right)$ to find the system in state \textbf{q} at time \textit{t} which corresponds to Eqs. (\ref{eq1A3}) and (\ref{eq2A3}), namely
\begin{eqnarray}
\nonumber
\frac{\partial W}{\partial t}\left( \mathbf{q},t \right)&=&\frac{1}{2{{\tau }_{c}}}\nabla \cdot \left[ \nabla W\left( \mathbf{q},t \right)+\beta W\left( \mathbf{q},t \right)\nabla {{V}_{T}}\left( \mathbf{q},t \right) \right]\\
&=&{{L}_{FP}}\left( \mathbf{q},t \right)W\left( \mathbf{q},t \right)
\label{eq3A3}
\end{eqnarray}
where $2{{\tau }_{c}}=\zeta /\left( kT \right)$ is the characteristic time of fluctuations, $\nabla $ is the del operator in \textbf{q} space, and ${{L}_{FP}}\left( \mathbf{q},t \right)$ is the Fokker-Planck operator. We notice that Eq. (\ref{eq3A3}) may also be written
\begin{eqnarray}
\frac{\partial W}{\partial t}\left( \mathbf{q},t \right)=\frac{1}{2{{\tau }_{c}}}\nabla \cdot \left\{ {{e}^{-\beta {{V}_{T}}\left( \mathbf{q},t \right)}}\nabla \left[ W\left( \mathbf{q},t \right){{e}^{\beta {{V}_{T}}\left( \mathbf{q},t \right)}} \right] \right\}
\label{eq4A3}
\end{eqnarray}
Now we use the transformation \cite{Kra40}
\begin{eqnarray}
\phi \left( \mathbf{q},t \right)=W\left( \mathbf{q},t \right){{e}^{\beta {{V}_{T}}\left( \mathbf{q},t \right)}}
\label{eq5A3}
\end{eqnarray}
so that  Eq. (\ref{eq4A3}) becomes
\begin{eqnarray}
\nonumber
\frac{\partial \phi }{\partial t}\left( \mathbf{q},t \right)-\beta \frac{\partial {{V}_{T}}}{\partial t}\left( \mathbf{q},t \right)\phi \left( \mathbf{q},t \right)&=&\frac{1}{2{{\tau }_{c}}}{{e}^{\beta {{V}_{T}}\left( \mathbf{q},t \right)}}\nabla \cdot \left\{ {{e}^{-\beta {{V}_{T}}\left( \mathbf{q},t \right)}}\nabla \phi \left( \mathbf{q},t \right) \right\}\\
&=&L_{FP}^{\dagger }\left( \mathbf{q},t \right)\phi \left( \mathbf{q},t \right)
\label{eq6A3}
\end{eqnarray}
where $L_{FP}^{\dagger }\left( \mathbf{q},t \right)$ is the adjoint Fokker-Planck operator \cite{Ris89}.

Next, we make the first approximation in our derivation, namely, we assume that the time variation of ${{V}_{T}}$ is small with respect to that of \textit{W}. If the time dependence of ${{V}_{T}}$ is contained in, say, the application of a time-varying uniform AC field only, this implies immediately that neglecting the second term in the left hand side of Eq. (\ref{eq6A3}) means that \textit{W} is near its equilibrium value, so restricting further calculations to low frequencies, $\omega {{\tau }_{c}}<<1$ (quasi-stationary condition). Hence, Eq. (\ref{eq6A3}) now reads
\begin{eqnarray}
\frac{\partial \phi }{\partial t}\left( \mathbf{q},t \right)\approx L_{FP}^{\dagger }\left( \mathbf{q},t \right)\phi \left( \mathbf{q},t \right)
\label{eq7A3}
\end{eqnarray}
Now, the interpretation of the Fokker-Planck equation (\ref{eq3A3}) (or equally well the Langevin equations (\ref{eq1A3})) with time-dependent potential in terms of usual population equations with time-dependent rate coefficients has a meaning, since now Eq. (\ref{eq5A3}) means detailed balancing. The polarization of an assembly of noninteracting DH in the direction of the applied field may then be defined as
\begin{eqnarray}
P\left( t \right)={{\rho }_{0}}{{\mu }_{d}}\int{\cos \vartheta \left( \mathbf{q} \right)W\left( \mathbf{q},t \right)d\mathbf{q}}
\label {eq8A3}
\end{eqnarray}
where $\rho_0$ is the number of DH per unit volume, and $\vartheta \left( \mathbf{q} \right)$ is the angle a DH dipole makes with the externally applied electric field. Because of the double-well structure of the interaction potential, we may equally well write Eq. (\ref{eq8A3}) 
\begin{eqnarray}
P\left( t \right)={{\rho }_{0}}{{\mu}_{d}} \left[ \int\limits_{well\,A}{\cos \vartheta \left( \mathbf{q} \right)W\left( \mathbf{q},t \right)d\mathbf{q}}+\int\limits_{well\,B}{\cos \vartheta \left( \mathbf{q} \right)W\left( \mathbf{q},t \right)d\mathbf{q}} \right]
\label {eq9A3}
\end{eqnarray}
Now, it is known from the Kramers theory of chemical reaction rates \cite{Kra40} that at sufficiently large energy barriers, most of the contributions of the integrands come from the minima of the wells, therefore we have
\begin{eqnarray}
P\left( t \right)\approx {{\rho }_{0}}{{\mu }_{d}}\left[ \cos \vartheta \left( {{\mathbf{q}}_{A}} \right)\int\limits_{well\,A}{W\left( \mathbf{q},t \right)d\mathbf{q}}+\cos \vartheta \left( {{\mathbf{q}}_{B}} \right)\int\limits_{well\,B}{W\left( \mathbf{q},t \right)d\mathbf{q}} \right]
\label{eq10A3}
\end{eqnarray}
Now, the integrals represent the relative populations ${{x}_{i}}\left( t \right)={{n}_{i}}\left( t \right)/N,\,\,\,i=A,B$ in each well (we assume that $W\left( \mathbf{q},t \right)$ is normalized to unity), where ${{n}_{i}}\left( t \right)$ is the number of DH states in well \textit{i}, and \textit{N} the total number of DH. At any time \textit{t}, we have the conservation law 
\begin{eqnarray}
{{x}_{A}}\left( t \right)+{{x}_{B}}\left( t \right)=1
\label{eq11A3}
\end{eqnarray}
Thus, Eq. (\ref{eq10A3}) reads
\begin{eqnarray}
P\left( t \right)\approx {{\rho }_{0}}\mu_d \left[ \cos \vartheta \left( {{\mathbf{q}}_{A}} \right){{x}_{A}}\left( t \right)+\cos \vartheta \left( {{\mathbf{q}}_{B}} \right){{x}_{B}}\left( t \right) \right]
\label{eq12A3}
\end{eqnarray}
We assume now for simplicity that $\vartheta \left( {{\mathbf{q}}_{B}} \right)=\pi -\vartheta \left( {{\mathbf{q}}_{A}} \right)$, so that
\begin{eqnarray}
P\left( t \right)\approx {{\rho }_{0}}\mu_d \cos \vartheta \left( {{\mathbf{q}}_{A}} \right)\left[ {{x}_{A}}\left( t \right)-{{x}_{B}}\left( t \right) \right]
\label{eq13A3}
\end{eqnarray}
Finally, since ${{\rho }_{0}}=N/V$ where \textit{V} is the volume of the polar substance made of DH only, we obtain
\begin{eqnarray}
P\left( t \right)\approx \frac{\mu_d \cos \vartheta \left( {{\mathbf{q}}_{A}} \right)}{N{{\upsilon }_{DH}}}\left[ {{n}_{A}}\left( t \right)-{{n}_{B}}\left( t \right) \right]
\label{eq14A3}
\end{eqnarray}
where ${{\upsilon }_{DH}}$ is the volume of a DH. This is the definition of the polarization in the Toy model.
\newline
In order to determine the polarization (\ref{eq14A3}), we need to calculate the dynamics of ${{n}_{i}}\left( t \right)$. From the conservation law -Eq. (\ref{eq11A3})-, we have
\begin{eqnarray}
{{\dot{x}}_{A}}\left( t \right)=-{{\dot{x}}_{B}}\left( t \right)=\int\limits_{well\,A}{\frac{\partial W}{\partial t}\left( \mathbf{q},t \right)d\mathbf{q}}
\label{eq15A3}
\end{eqnarray}
By using the Fokker-Planck equation (\ref{eq4A3}) and limiting well \textit{A} to a closed generalized bounding surface constituting the saddle region $\partial A$, we have by Gauss's theorem
\begin{eqnarray}
{{\dot{x}}_{A}}\left( t \right)=-{{\dot{x}}_{B}}\left( t \right)=\frac{1}{2{{\tau }_{c}}}\oint\limits_{\partial A}{{{e}^{-\beta {{V}_{T}}\left( \mathbf{q},t \right)}}\nabla \phi \left( \mathbf{q},t \right)\cdot {{\mathbf{\nu }}_{\mathbf{q}}}d{{S}_{\mathbf{q}}}}
\label{eq16A3}
\end{eqnarray}
where $\mathbf{\nu_q}$ is the outward normal to the bounding surface and $dS_{\mathbf{q}}$ is a generalized surface element of the bounding surface, and where we have used Eq. (\ref{eq5A3}). Now, we follow closely Coffey et al. \cite{Cof01} and introduce the crossover function $\Delta \left( \mathbf{q},t \right)$ via the equation
\begin{eqnarray}
\phi \left( \mathbf{q},t \right)={{\phi }_{A}}\left( t \right)+\left[ {{\phi }_{B}}\left( t \right)-{{\phi }_{A}}\left( t \right) \right]\Delta \left( \mathbf{q},t \right)
\label{eq17A3}
\end{eqnarray}
where $\Delta \left( \mathbf{q},t \right)=0$ if $\mathbf{q}\in well\,A$ while $\Delta \left( \mathbf{q},t \right)=1$ if $\mathbf{q}\in well\,B$ and exhibits strong gradients in the saddle region $\partial A$ allowing the crossing from \textit{A} to \textit{B} (and vice-versa) by thermally activated escape. By combining Eqs. (\ref{eq16A3}) and (\ref{eq17A3}), we have immediately
\begin{eqnarray}
{{\dot{x}}_{A}}\left( t \right)=-{{\dot{x}}_{B}}\left( t \right)=\frac{{{\phi }_{B}}\left( t \right)-{{\phi }_{A}}\left( t \right)}{2{{\tau }_{c}}}\oint\limits_{\partial A}{{{e}^{-\beta {{V}_{T}}\left( \mathbf{q},t \right)}}\nabla \left[ \Delta \left( \mathbf{q},t \right) \right]\cdot {{\mathbf{\nu }}_{\mathbf{q}}}d{{S}_{\mathbf{q}}}}
\label{eq18A3}
\end{eqnarray}
Now,
\begin{eqnarray}
{{x}_{i}}\left( t \right)={{\phi }_{i}}\left( t \right)x_{i}^{s}\left( t \right),\,\,\,x_{i}^{s}\left( t \right)=\int\limits_{well\,i}{{{W}_{s}}\left( \mathbf{q},t \right)d\mathbf{q}}
\label{eq19A3}
\end{eqnarray}
where ${{W}_{s}}\left( \mathbf{q},t \right)$ is a normalized solution of the Fokker-Planck equation
\begin{eqnarray}
{{L}_{FP}}\left( \mathbf{q},t \right){{W}_{s}}\left( \mathbf{q},t \right)=2{{\tau }_{c}}\frac{\partial {{W}_{s}}}{\partial t}\left( \mathbf{q},t \right)\approx 0
\label{eq20A3}
\end{eqnarray}
because the frequencies we are concerned with are very small with respect to the inverse thermal fluctuation time ${{\tau }_{c}}$ and because the time-dependent part of the potential ${{V}_{T}}$ is much smaller than other terms in it at any time. We have
\begin{eqnarray}
x_{A}^{s}\left( t \right)+x_{B}^{s}\left( t \right)=1
\label{eq21A3}
\end{eqnarray}
Using Eqs. (\ref{eq19A3}) and (\ref{eq21A3}), we may easily show that \cite{Cof01}
\begin{eqnarray}
{{\phi }_{B}}\left( t \right)-{{\phi }_{A}}\left( t \right)=\left( \frac{1}{x_{A}^{s}\left( t \right)}+\frac{1}{x_{B}^{s}\left( t \right)} \right)\left[ {{x}_{B}}\left( t \right)x_{A}^{s}\left( t \right)-{{x}_{A}}\left( t \right)x_{B}^{s}\left( t \right) \right]
\label{eq22A3}
\end{eqnarray}
By combining Eqs. (\ref{eq16A3}) and (\ref{eq22A3}), we readily obtain
\begin{eqnarray}
{{\dot{x}}_{A}}\left( t \right)=-{{\dot{x}}_{B}}\left( t \right)=\Gamma \left( t \right)\left( {{x}_{B}}\left( t \right)x_{A}^{s}\left( t \right)-{{x}_{A}}\left( t \right)x_{B}^{s}\left( t \right) \right)
\label{eq23A3}
\end{eqnarray}
where the overall time-dependent escape rate $\Gamma \left( t \right)$ is given by \cite{Cof01}
\begin{eqnarray}
\Gamma \left( t \right)=\frac{1}{2{{\tau }_{c}}}\left( \frac{1}{x_{A}^{s}\left( t \right)}+\frac{1}{x_{B}^{s}\left( t \right)} \right)\oint\limits_{\partial A}{{{e}^{-\beta {{V}_{T}}\left( \mathbf{q},t \right)}}\nabla \phi \left( \mathbf{q},t \right)\cdot {{\mathbf{\nu }}_{\mathbf{q}}}d{{S}_{\mathbf{q}}}}
\label{eq24A3}
\end{eqnarray}
Finally, by setting
\begin{eqnarray}
{{\Pi }_{AB}}\left( t \right)=\Gamma \left( t \right)x_{B}^{s}\left( t \right),\,\,\,\,\,{{\Pi }_{BA}}\left( t \right)=\Gamma \left( t \right)x_{A}^{s}\left( t \right)
\label{eq25A3}
\end{eqnarray}
we arrive at the population equations
\begin{eqnarray}
{{\dot{n}}_{A}}\left( t \right)=-{{\dot{n}}_{B}}\left( t \right)=-{{\Pi }_{AB}}\left( t \right){{n}_{A}}\left( t \right)+{{\Pi }_{BA}}\left( t \right){{n}_{B}}\left( t \right)
\label{eq26A3}
\end{eqnarray}
The obtaining of a more explicit formula for the various rates involved in Eq. (\ref{eq25A3}) is not possible, due to the impossibility to calculate the surface integral in Eq. (\ref{eq24A3}) explicitly, in turn due to the fact that $V_{T}$ is not known explicitly. Then, the rates in Eqs.  (\ref{eq25A3}) and (\ref{eq26A3}) are estimated using Arrhenius's formula. All subsequent derivations regarding the Toy model of Ladieu et al. \cite{Lad12} follow immediately and will not be repeated here due to lack of room and straightforward but laborious algebra.


\begin{thebibliography}{99.}
\bibitem{note9} It would be worth to develop further the MCT analysis of \cite{Tar10} to obtain detailed analytical predictions on the phases.
\bibitem{note12} The growth of $\chi_3$ and $\chi_5$ can be induced by purely dynamical correlation, as it is the case in MCT theory and in the MCT regime.
However, in the activated dynamics regime, the only theory that can explain a growth of $\chi_3$ and $\chi_5$ are the ones in which
dynamical correlations are driven by static (``point-to-set'') correlations.
\end{thebibliography}
\end{document}